\documentclass[a4paper,fleqn,usenatbib]{mnras}

\usepackage{newtxtext,newtxmath}
\usepackage[T1]{fontenc}
\usepackage{ae,aecompl}

\usepackage{graphicx}	% Including figure files
\usepackage{amsmath}	% Advanced maths commands
\usepackage{amssymb}	% Extra maths symbols

\usepackage{amsfonts}
\usepackage[colorinlistoftodos]{todonotes}
\usepackage{hyperref}
\usepackage{hf-tikz}
\usepackage{graphicx}
\usepackage{color}
\usepackage{booktabs}
\usepackage{lineno}
\usepackage{enumerate}
\usepackage[breakable, theorems, skins]{tcolorbox}
\DeclareRobustCommand{\mybox}[2][gray!20]{%
\begin{tcolorbox}[   %% Adjust the following parameters at will.
        breakable,
        left=0pt,
        right=0pt,
        top=0pt,
        bottom=0pt,
        colback=#1,
        colframe=#1,
        width=\dimexpr0.95\columnwidth\relax, 
        enlarge left by=0mm,
        boxsep=5pt,
        arc=0pt,outer arc=0pt,
        ]
        #2
\end{tcolorbox}
}

\title[Stability of exosystems]{Stability of exoplanetary systems retrieved from scalar time series}

\author[T. Kov\'acs]{
T. Kov\'acs,$^{1}$\thanks{E-mail: tkovacs@general.elte.hu (TK)}
\\
$^{1}$Institute of Physics, E\"otv\"os University, 1117 Budapest, P\'azm\'any P. s. 1A, Hungary\\
}

\date{Accepted XXX. Received YYY; in original form ZZZ}

\pubyear{2019}

\begin{document}
\label{firstpage}
\pagerange{\pageref{firstpage}--\pageref{lastpage}}
\maketitle

% Abstract of the paper
\begin{abstract}
%Dynamical stability is one of the main characterestics of an extrasolar
%planetary system. The increasing number of discovered multi-planet
%systems and their dynamical diversity implies efficient methods to
%predict planetary motion.

We propose a novel method applied to extrasolar planetary dynamics to describe the system stability. The observations in this field serve the measurements mainly of radial velocity, transit time, and/or celestial position. These scalar time series are used to build up the high-dimensional phase space trajectory representing the dynamical evolution of planetary motion. The framework of nonlinear time series analysis and Poincar\'e recurrences allows us to transform the obtained univariate signals into complex networks whose topology carries the dynamical properties of the underlying system. The network-based analysis is able to distinguish the regular and chaotic behaviour not only for synthetic inputs but also for noisy and irregularly sampled real world observations. The proposed scheme does not require neither n-body integration nor best fitting planetary model to perform the stability investigation, therefore, the computation time can be reduced drastically compared to those of the standard numerical methods. 
  
\end{abstract}

% Select between one and six entries from the list of approved keywords.
% Don't make up new ones.

%%%%%%%%%%%%%%%%%%%%%%%%%%%%%%%%%%%%%%%%%%%%%%%%%%

%%%%%%%%%%%%%%%%% BODY OF PAPER %%%%%%%%%%%%%%%%%%

\section{Introduction}

Since the gravitationally interacting, and therefore non-Keplerian,
few-body dynamics can be very complex \citep{Fabrycky2010}, it makes a great deal to
investigate the stability of exoplanetary systems containing at least
two planets \citep{Rivera2001,Armstrong2015,Batygin2015,Gozdziewski2016,Panichi2018}.
Dynamical modeling of multiple planetary systems requires precise
initial conditions and system parameters in order to 
perform reliable n-body numerical integration. The most common methods to
constrain the planetary masses and orbital elements are radial
velocity (RV) measurements \citep{Laughlin2001,Rivera2001,Tan2013,Nelson2014} and transit timing observations, especially their variation (TTVs) \citep{Agol2005,Holman2005}. 
RV \citep{Laughlin2006,Pal2010} and TTV \citep{Nesvorny2008,Nesvorny2010,Veras2011,Deck2014,Panichi2018} data sets together provide the input of sophisticated but fairly
time consuming statistical methods (e.g. \citet{Foreman-Mackey2013}), which can serve the best fitting
models in high dimensional parameter space (with a certain confidence
level) of the underlying planetary dynamics. From the obtained
parameters one then can predict the long-term stability of the
system.

In addition, low-order analytic and semi-analytic inversion methods often fail close to dynamical degeneracy such as mean motion resonances (MMRs) that are frequent among the
known exoplanetary systems \citep{Fabrycky2014}. This shortcoming has been recently avoided by analytic models of transit timing variation for moderate eccentricities and certain range of masses \citep{Agol2015,Deck2016,Hadden2016}. These methods are suitable to resolve the degeneracy between planetary masses and eccentricities close to low order MMRs.

Nonlinear dynamical systems may often produce 
chaotic behaviour which means that they are extremely sensitive even
for a small perturbation of the initial conditions \citep{Ott2002,Tel2006}. This fact makes
the application of the above mentioned methods even more problematic or limited to certain conditions.
Describing the dynamics of a known deterministic nonlinear dynamical
system, when the equations of motion are known, means basically exploring
the phase space patterns.
This is, however, not the case in the exoplanetary research. In practice,
we measure only one signal of the underlying dynamics, and have to
obtain the system's behavior from this scalar time series. 
The question whether the dynamical invariants (such as Lyapunov
exponents or other measures of irregularity) of a particular system
can be recovered from a single variable data set is, therefore, extremely important.

In this work a widely used nonlinear data analysis method is proposed which is
based on the fundamental 
theorem known as Poincar\'e recurrence \citep{Saussol2002}. In short, any conservative dynamical system recurs sooner or later
to its former states in phase space. Recently, in their report
\citet{Marwan2007} showed how the visual interpretation of recurrences, the so-called recurrence plots (RPs) \citep{Eckmann1987}, are quantitatively related to the characteristic of a dynamical system.
A generalization of RPs to network representation  \citep{Donner2010,Donner2011} widely extended the adaptability of recurrences.
Application of recurrence network analysis to
cutting-edge measurements in exoplanetary research provides a robust
and novel method to investigate the stability. Besides the numerical
integration of the best fitting planetary models we propose a
complementary study to describe the system stability without having
any knowledge about the parameters and initial conditions. 

The paper is organized as follows. In Section~\ref{sec:model}, a case
study of a two-planet system is presented as a basic dynamical model.
The data analysis method is thoroughly
described in Section~\ref{sec:method}. Section~\ref{sec:data} examines the application
to real exoplanetary data. The main conclusions are drawn in the final section.

\section{The model system}
\label{sec:model}

At this point a simple dynamical model is introduced that will
provide synthetic time series (RV, TTV, and astrometry positions) in
order to establish and test the recurrence-based data 
analysis method in Section~\ref{sec:method}. The Sun, Jupiter,
Saturn (SJS) subset of our own Solar System has been chosen to be the
basis of the stability analysis. The initial orbital elements of the
planets are set to be those at the epoch of J2000 (JD 2451545.0), see
Table A.2. in \cite{SSD2000}.

\begin{center}
\begin{figure}
\includegraphics[angle=0,width=\columnwidth]{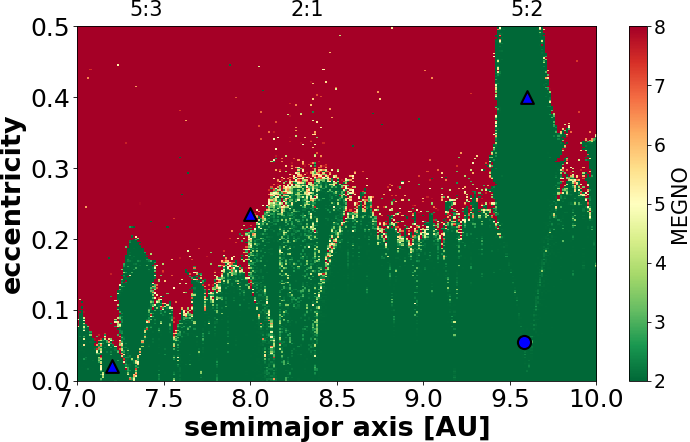}
\caption{Stability map of Saturn ($a_{\mathrm{Saturn}},e_{\mathrm{Saturn}}$) in SJS
  system. The stability index MEGNO is color coded according to the color-bar and maximized by 8. The blue triangles indicate the test data series with different dynamical behaviour as a base of analysis. From left to right: (7.2,0.02) regular non-resonant, (8.0,0.235) chaotic, (9.6,0.4) regular resonant. The blue circle depicts the Saturn's actual position in the $a-e$ parameter plane. The most prominent MMRs are also marked at the top of the panel.}
\label{fig:megno}
\end{figure}
\end{center}

The well-known semimajor axis--eccentricity ($a,e$) stability map of the model
system is shown in Figure~\ref{fig:megno}. The map is obtained as follows. The
barycentric coordinates of the Sun and the two massive planets were
integrated by using the \texttt{REBOUND-WHFAST} routine \citep{Rein2015} over 1000 periods of the
inner body, i.e. Jupiter. In 
addition, Saturn's orbital elements cover a grid of 250x250 initial
conditions. That is $a_{\mathrm{Saturn}}=[7.5:10],\; \delta a_{\mathrm{Saturn}}=0.01$ and $e_{\mathrm{Saturn}}=[0:0.5],\;
\delta e_{\mathrm{Saturn}}=0.002.$ Therefore, in these plots, the
semimajor axis and the eccentricity always refer to Saturn's initial
orbital elements in the rest of the paper. In order to characterize the system stability the chaos indicator
MEGNO (Mean Exponential Growth of Nearby Orbits, \citet{Cincotta2000}) 
is calculated for every set of initial conditions $(a_{\mathrm{Saturn}},e_{\mathrm{Saturn}}).$ The system
is regular for values about 2 (green) while chaotic for larger MEGNO
values (red). It has to be emphasized that most of the chaotic
trajectories lead to the disruption of the system. As a result, the
integration stops when one of the planets escapes the system. In this
case the value of MEGNO is set to be 8. The wide green bands penetrating into the chaotic domain
correspond to certain MMRs. The rightmost resonance around $a=9.6$ is
the 5:2 commensurability between Jupiter and Saturn.

\begin{center}
  \begin{figure}
    \includegraphics[angle=0,width=\columnwidth]{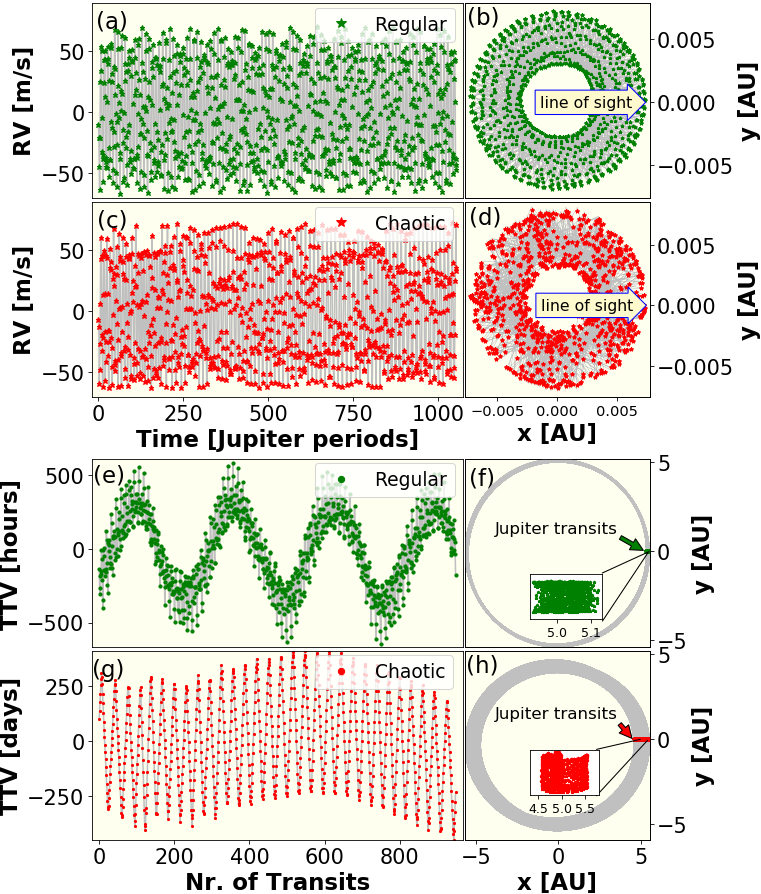}    
\caption{Numerically generated times series for the SJS system. (a) and (c): Radial velocity (RV) of
  the Sun along the x-axis. (b) and (d): Celestial positions of the Sun in the
  x-y plane. (e) and (g): Transit timing variation (TTV) of Jupiter. The green and red dots belong to the initial
  conditions (7.2,0.02) and (8.0,0.235) marked by blue triangles in Fig~\ref{fig:megno}, regular and chaotic respectively. The line of sight is the (positive) x-axis. (f) and (h): Transit positions of Jupiter. The insets highlight the radial ''spread'' (in AU) of transit positions in the x-y plane.}
\label{fig:RV_TTV}
\end{figure}
\end{center}

In order to apply the recurrence-based technique to a scalar time series,
all numerically computed phase space trajectories are stored and re-used later. 
This allows one to generate synthetic radial velocity (measured along
the x-axis as the line of sight), star position ((x,y) co-ordinates
are stored as face on view), and transit time data sets (viewed again
from the x-axis) imitating the ideal (noise-free and equidistant)
astronomical observations. Some examples are demonstrated in
Figure~\ref{fig:RV_TTV}. The upper four panels show the RV (a,c) and
celestial positions (b,d) of the Sun containing 950 data points. The
remaining part of the Figure deals with Jupiter's TTV signal (e,g) and
its mid-transit position (f,h) with a precision of ~10$^{-5}$ day.

The integration time for the RV data is ca. 1050 orbital periods of Jupiter which means that the
sampling frequency is less than the mean motion of the inner
planet. TTV signals carry 950 data points similar to those of RV data
in order to have the same length of time series in the analysis phase. 
From now on, these data represent the measurements subject to
examination. 

The length of the time series is chosen to cover a realistic time
frame, i.e around 1000 periods of Jupiter. It might seem to be too
long on human time scales but it is not uncommon in currently known,
especially tightly packed, exoplanetary systems. For example, 1000
orbital periods of the inner planets (b-e) in the TRAPPIST-1 system
corresponds to 1500-6000 days, ca. 4-16 years, the same number of orbital
revolution for the Kepler-18 system requires 11-20 years, and even
less for Kepler-412. The length of 950 data points
can be achieved with the upcoming surveys of the near
future. 

Here, we take the opportunity and explain a hidden phenomenon between
synthetic RV and TTV time series which is not crucial for the analysis
but is worth clarifying. Most of the calculations are done for 1050
Jupiter periods in order to obtain the RV signal of the Sun. This data
set is then sampled by 950 points equidistantly given a time series to be
analysed. This means that every single RV measurement is considered with equal length given above. In contrast,
obtaining the TTV signals we set the length of the data to 950 Jupiter
transits. Clearly, the appearance of transits depends on the dynamics,
as well as, on the initial configuration of the system. That is, as
the orbital elements change in time the 950 TTVs can be obtained
sooner or later and, therefore, the time series being analysed might, and
actually do, somewhat differ in their length.

\section{Methods and techniques}
\label{sec:method}

\begin{center}
\begin{figure}
\includegraphics[angle=0,width=\columnwidth]{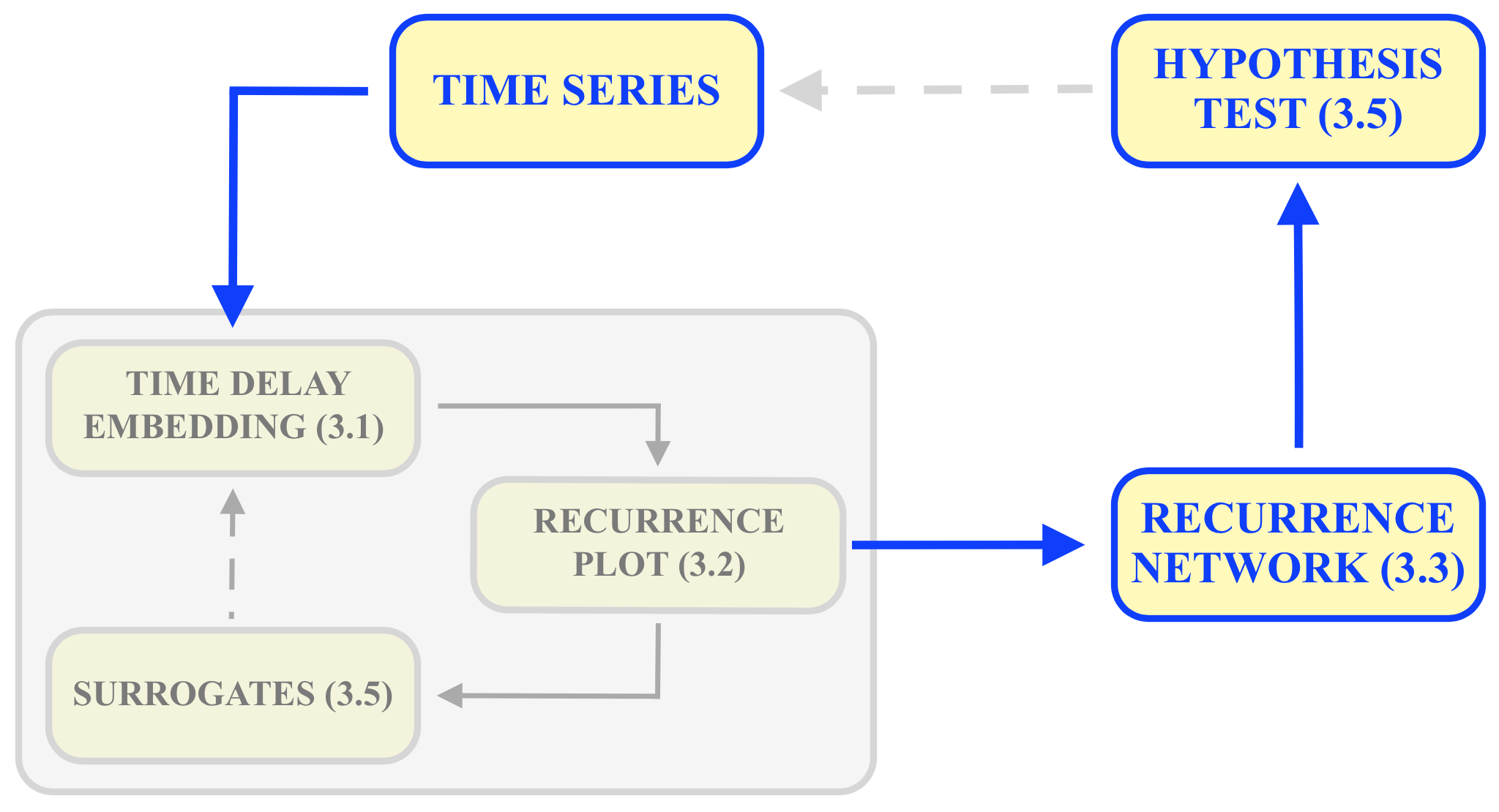}
\caption{Schematic work flow of the nonlinear time series
  analysis. Blue boxes are input and output, gray part contains the
  technical infrastructure.}
\label{fig:work-flow}
\end{figure}
\end{center}

In this section the data analysis method based on phase space
recurrences is discussed in detail. Synthetic time series are generated as
demonstrated in the previous section. In some cases to have a more realistic
scenario noise is added to data. Where this applies, a precise
quantitative description is given in the text. This section is divided
into five subsections according to the schematic work flow in
Figure~\ref{fig:work-flow}. All the numbered boxes are relevant part
of the analysis and the results from one box in the work flow are then
used in the next box according to the blue arrows (starting with a
time series in counterclockwise direction). Subsection~\ref{sec:reconstruction} gives
an overview about the phase space reconstruction. The second
subsection is devoted to recurrence plots. The link between RPs and recurrence networks is examined in
Subsection~\ref{sec:RN}. Noisy and unevenly sampled data analysis is
introduced in Subsection~\ref{sec:noise_gap}. Finally, the role of
surrogate data analysis and hypothesis testing is demonstrated. This
sequence of topics can be thought of as a work flow of the whole
procedure. The reader might feel that this
part of the paper is somewhat lengthy, we think, however, that the
details presented here are pedagogically necessary to demonstrate how the method
works in general before applying it to real data.

\subsection{Phase space reconstruction}
\label{sec:reconstruction}

The dynamical analysis of a particular system requires the evolution
of phase space trajectories. Since in an experimental setting the observer
records the signals in time domain, which means not all relevant
components of the state vector is known, a reconstruction of multidimensional
information in an artificial phase space is needed. This is possible when the following assumption
holds \citep{Semmlow}: the variable corresponding to the observable affects the other state space
variables, i.e. the variables governing the system's dynamics are
coupled. In other words, all hidden variables in the system make some
contribution to the measured signal. In this case, a  
recovery of the phase space trajectory can be done by using an embedding
theorem \citep{Takens1981,Mane1981,Packard1980}. In what follows, the method of time delay reconstruction is interpreted. 

\subsubsection{Time delay embedding}
\label{sec:timedelay}

Time series $x(t_i)$ is a sequence of $i=1,\dots,\;n$ scalar measurements of some
quantity ($x$) depending on the state of the system ($\mathbf{s}$) taken at discrete times ($\Delta t$) 
\begin{equation}
x(t_i) = x(\mathbf{s}(n\Delta t)).
\end{equation}

The reconstructed $m$ dimensional vector reads \citep{Kantz2003} then
\begin{equation}
x(t_i) \rightarrow \mathbf{x}_N = \{x(t_i-(m-1)\tau), x(t_i-(m-2)\tau),\dots, x(t_i-\tau), x(t_i))\},
\label{eq:tde}
\end{equation}
where $i=1\dots n$ is the length of the original signal, $m$ is the dimension
into which the reconstructed vector is embedded, the delay, $\tau,$ is
the time difference between adjacent components of $\mathbf{x}.$

That is, the method of delays allows one to transform an observed
scalar time series into a higher dimensional representation in the
embedding space. The advantage of this technique is that it provides a
one-to-one image of the original (multi-dimensional) state vector $\mathbf{s}$.

After the delay reconstruction the length of $\mathbf{x}$ is
reduced to $N=n-(m-1)\tau.$ That is, the components of the reconstructed
vector are segments of the original 1D signal delayed by $\tau.$ Thus,
$\mathbf{x}$ is an $m\times [n-(m-1)]\tau$ matrix. For clarity consider the example in Appendix~\ref{app:delay_example}.

\begin{center}
\begin{figure*}
\includegraphics[angle=0,scale=.4]{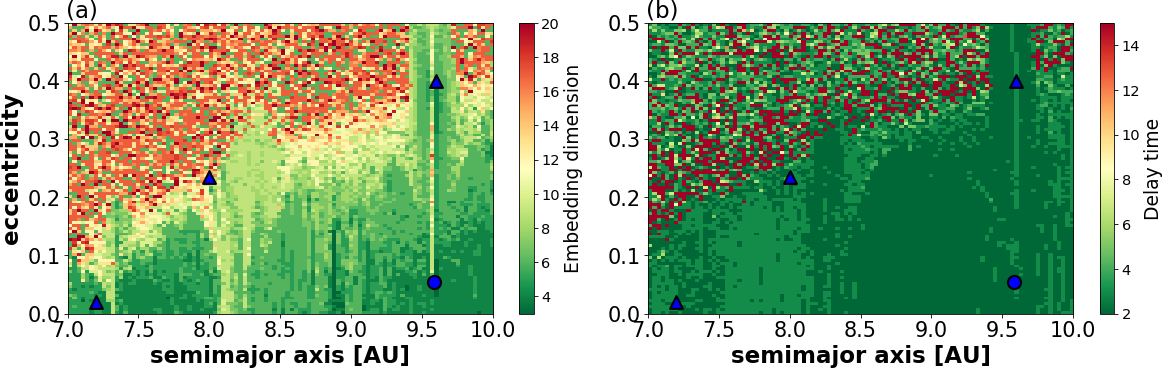}
\caption{Embedding dimension (a) and time delay (b) dependence for ($a_{\mathrm{Saturn}},e_{\mathrm{Saturn}}$) initial conditions similar as in Figure~\ref{fig:megno}. The resolution is 100x100 points each directions. The synthetic data sets include 950 measurements of Sun's radial velocity during the 1050 Jupiter-periods. Marked points correspond to different dynamical regimes (blue triangles) and the Saturn's actual position (circle) in the ($a,e$) plane, see also Figure~\ref{fig:megno}. Similar plots can also be constructed from the TTVs and celestial positions, not presented here.}
\label{fig:emb_tau}
\end{figure*}
\end{center}

Takens' embedding theorem
\citep{Takens1981} states that the dynamical properties of a $d$ dimensional
attractor can be reconstructed if $m>2d$ no matter how large the
original dimension of the true state space is\footnote{It has been shown that fewer dimensions are sufficient for measured data.}. Takens also showed that the time lag can have arbitrary value, however, there are practically relevant values.

It should be noted that the reconstructed trajectory
is not identical to that one we would have from numerical integration,
i.e. when all the components are known. It might differ in shape but
preserves the mathematical properties such as topology and Lyapunov exponents. 

At this point a natural question might arise. How to find the
right embedding dimension and time delay? Since there are many widely
accepted methods to choose appropriate values \citep{Kantz2003,Ma2006} here only those are
reported that were used in the current analysis. The embedding dimension $m$
can be obtained by using the concept of false nearest neighbours. The primary goal of this procedure is to find neighbours in embedding space that become not neighbours anymore because the temporal evolution, called false nearest neighbours (FNN). For a 
detailed description see \citet{Kantz2003} and references therein. One method for
estimating the time delay $\tau$ is to find the first minimum of the
mutual information function (MIF) \citep{Semmlow}. MIF can be considered as a
nonlinear analog of the autocorrelation function and, therefore, it is more
appropriate in this analysis. More detailed expalantation and examples of FNN and MIF can be found in Appendix~\ref{app:delay_example}. The \texttt{TISEAN}\footnote{\url{https://www.pks.mpg.de/~tisean/Tisean_3.0.1/}} software package \citep{Hegger1999} has been used to calculate the delay parameters through the whole analysis.

At the end of this subsection the parameters of the delay
reconstruction ($m,\tau$) in SJS system are presented. The previously introduced grid of
initial conditions $(a_{\text{Saturn}},
e_{\text{Saturn}}$) is used, though, the embedding dimension and
the time delay is plotted instead of the stability index
MEGNO, see Figure~\ref{fig:emb_tau}. Each grid point in the ($a_{\text{Saturn}},e_{\text{Saturn}}$) parameter plane represents a single time series, Sun's synthetic RV, that has been obtained as interpreted in Figure~\ref{fig:RV_TTV}.

Both maps are similar to those showing the stability index MEGNO in
Fig.~\ref{fig:megno}. However, the heat map of embedding dimension,
Figure~\ref{fig:emb_tau}(a), catches more details than the time delay
map (Figure~\ref{fig:emb_tau}(b)). We emphasis here that neither $m$ nor $\tau$ do not represent strict dynamical properties of the underlying system. These two parameters are, nevertheless,  the backbone of the phase space trajectory reconstruction. Indeed, one might expect some relation between time delay embedding and system dynamics. This can be observed evidently in case of
embedding dimension. The chaotic and regular parts are clearly distinguishable although the border is not as sharp as in case of direct calculation of MEGNO. As a matter of fact,  the more irregular the dynamics, the less the correlation between its elements.  In other words, a motion with increased random elements (i.e. stronger instability) may require larger embedding dimension than ordered motion \citep{Stergiou}. In case of a strong gravitational interaction, one of the planets can
be ejected from the system. If this happens before the integration is
over, $m$ and $\tau$ are set to be a large number, 100,
and the corresponding initial condition is red.

We can conclude from both panels that the optimally determined embedding parameters
carry the information rather accurately about the underlying
dynamics. Moreover, it is also demonstrated that the
requested time delay, $\tau,$ indicates a lesser degree of correlation in the
dynamics also in agreement with Takens theorem.

\subsection{Recurrence Plots}
\label{sec:RP_RQA}

Extracting meaningful information from a data set can be easy if
regular patterns of the observable appear in time domain. However,
when the time series is more complex and no simple rule for its time
dependence can be formulated, the representations of certain events might
help us to draw some conclusion about the dynamics.

Once the phase space trajectories are available, either the original or
the reconstructed one, these representations show up when the state vector
returns to a neighbourhood of a point that has already been visited. This phenomenon was first described by Poincar\'e and is called
as recurrence. The related recurrence time statistics became a key concept of dynamical system analysis across many disciplines. These recurrences in phase space can be easily
visualized by recurrence plots (RP) originally introduced by \citet{Eckmann1987}. 

An RP is a very simple method for measuring and visualizing
recurrences of a trajectory even in higher dimensions. It can be
represented by a two-dimensional matrix $\mathbf{R}$
\begin{equation}
R_{i,j}(\epsilon)=\Theta(\epsilon-||\mathbf{x}_i-\mathbf{x}_j||),\quad
i,j=1\dots N,
\label{eq:RP}
\end{equation}
where $N$ is the length of the (reconstructed) phase space trajectory, $\Theta(\cdot)$ is the
Heaviside step function, $\epsilon$ a tolerance parameter, and
$||\cdot||$ is a norm. The embedded delay vectors obtained from
measured points are $\mathbf{x}_i$ and $\mathbf{x}_j$ at different
time instant, $t_i$ and $t_j,$ respectively. If a trajectory at $t_j$
returns to an $\epsilon$ neighbourhood of a point where it was at $t_i$
($t_j>t_i$) then the corresponding matrix element is 1,
i.e. recurrence occurs, otherwise 0\footnote{The recurrence rate $RR$
  is the ratio of 1 to 0 elements in the recurrence matrix.}.
To be more precise

\begin{equation}
  R_{i,j}(\epsilon)=
  \left\{
    \begin{array}{@{}ll@{}}
      0, & \text{when}\ \epsilon < ||\mathbf{x}_i-\mathbf{x}_j||, \\
      1, & \text{when}\ \epsilon > ||\mathbf{x}_i-\mathbf{x}_j||.
    \end{array}\right.
\end{equation}
% https://tex.stackexchange.com/questions/9065/large-braces-for-specifying-values-of-variables-by-condition

\begin{center}
\begin{figure}
\includegraphics[angle=0,width=\columnwidth]{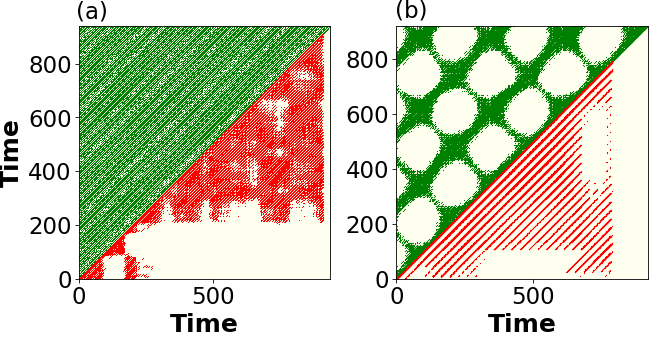}
\caption{Recurrence plots for two different trajectories in SJS system
  based on RV data sets (a) and Jupiter's TTV signals (b). Integration time: 1050 Jupiter
  periods with 950 measurements for RVs and 950 TTVs. The recurrence rate ($RR$) is fixed to
  0.1. The phase space distance is defined by the maximum norm.} 
\label{fig:rp}
\end{figure}
\end{center}

The matrix $\mathbf{R}$ is symmetric by definition. Plotting the
elements of the binary matrix with different colors, one can obtain the
RP. Figure~\ref{fig:rp} depicts four recurrence plots corresponding to
two initial conditions marked in Fig.~\ref{fig:megno} (blue triangles). In panel (a), the lower right triangle (red) depicts the RP corresponding to the reconstructed
trajectory from Sun's RV data originating from
$(a_{\text{Saturn}},e_{\text{Saturn}})=(8.0,0.235),$ the upper left (green) is for
$(a_{\text{Saturn}},e_{\text{Saturn}})=(7.2,0.02).$ Panel (b) shows the same obtained from the TTV signals of Jupiter.
One can see the differences between the regular (green) and chaotic (red)
dynamics. Basically, RPs have different patterns associated with 
different kind of time evolution of trajectories (periodic,
quasi-periodic, chaotic) \citep{Zou2007,Ngamga2012}. At the very basic level, dots
typically appear as diagonal line segments on RP. Periodic signals
yield non-interrupted lines, while chaotic dynamics results in a
pattern of diagonals much shorter than for periodic cycles.
For a profound review about different structures of RPs see Refs.~\citet{Marwan2004,Marwan2007}.

Equation~(\ref{eq:RP}) reveals that RP depends on the threshold
$\epsilon$ and norm $||\cdot||.$ In the literature there are many
examples how to select $\epsilon.$ Several rules of thumb can be found
such as (i) it should not exceed 10\% of the  maximum diameter of the
phase space, or (ii) $\epsilon$ is at least five times the STD of the
observed noise \citep{Stergiou,Semmlow}. It is clear, if the threshold varies, it results in different density of points in RP. 

An alternative approach is that we fix RR, i.e the ratio of 1
to 0, in $\mathbf{R}$ and choose a dynamic threshold that provides a
constant density of points in the RP. This practical choice avoids the
sparse structure of RP due to a low threshold, and helps to
make consistent analysis with constant density of points.

The commonly used norms for the same distance between two points are $L_1$-norm (Manhattan norm), $L_2$-norm
(Euclidean distance), and $L_{\infty}$-norm (Maximum norm). For the sake of clarity, definitions of different lengths in phase space are given in Appendix~\ref{app:norm}. To construct the RPs in this work dynamic threshold (with recurrence rate
4-10\%) and maximum norm is employed, see Equation~(\ref{eq:max_norm}).

\subsection{Recurrence Networks}
\label{sec:RN}

A natural way to analyse complex systems is using complex network
theory developed in the last two decades. The fact that nonlinear time
series analysis can be used effectively to study complex dynamics and
the successful application of networks in various fields stimulated
the demand of transforming time series into complex
networks. Recently, several different methods have been proposed
\citep{Xu2008,Shimada2008,Donner2010}. In many cases a network can be describe mathematically as a
graph $G(V,E)$ where $V=\{1,2,\dots N\}$ is a set of vertices and
$E\subseteq V\times V$ represents the edges between pairs of
vertices. In case of unweighted and undirected networks  a symmetric
$N\times N$ adjacency matrix ($A_{i,j}$) completely describes the
graph $G$ 
\begin{equation}
  A_{i,j}=
  \left\{
    \begin{array}{@{}ll@{}}
      1, & \text{when}\ i,j \in E, \\
      0, & \text{otherwise}\ .
    \end{array}\right.
\end{equation}

Those not familiar with networks are encouraged to read
Appendix~\ref{app:network} before going to applications.

In this work only proximity networks, a subclass of networks which are
directly related to Poincar\'e recurrences, will be used.  
An $\epsilon$-recurrence network (RN) consists of vertices formed by
reconstructed state vector in the phase space ($\mathbf{x}_N,$ see Section~\ref{sec:timedelay})
connected by edges to other vertices in their $\epsilon$
neighborhood. This definition allows to re-interpret the recurrence
matrix ($R_{i,j}$) of an RP in the following way 
\begin{equation}
A_{i,j}=A_{i,j}(\epsilon)=R_{i,j}(\epsilon)-\delta_{ij},
\label{eq:adjmat}
\end{equation}
where $A_{i,j}$ is the adjacency matrix of a complex network embedded
in a phase space, Kronecker delta ($\delta_{ij}$) avoids the
self-loops in the graph. The matrix $\mathbf{A}$ carries the symmetry
properties of matrix $\mathbf{R},$ consequently RN is a graph with no
self-loops and multiple edges. Equation~(\ref{eq:adjmat}) illuminates
an alternative view of RPs, namely, the joint proximity observations
in phase space can be represented as links in complex
networks. Moreover, $A_{i,j}$ describes random geometric graphs where
the vertices are located in a metric space such as the reconstructed
phase space. The quantitative description of RNs, to be discussed
later, provides relevant geometric information about the underlying
dynamics. Furthermore, recurrence network method is not based on temporal correlations or explicit time
ordering \citep{Donner2011}. This property of RNs will be
extremely useful in case of irregularly sampled data sets.

\begin{center}
\begin{figure*}
\includegraphics[angle=0,scale=.4]{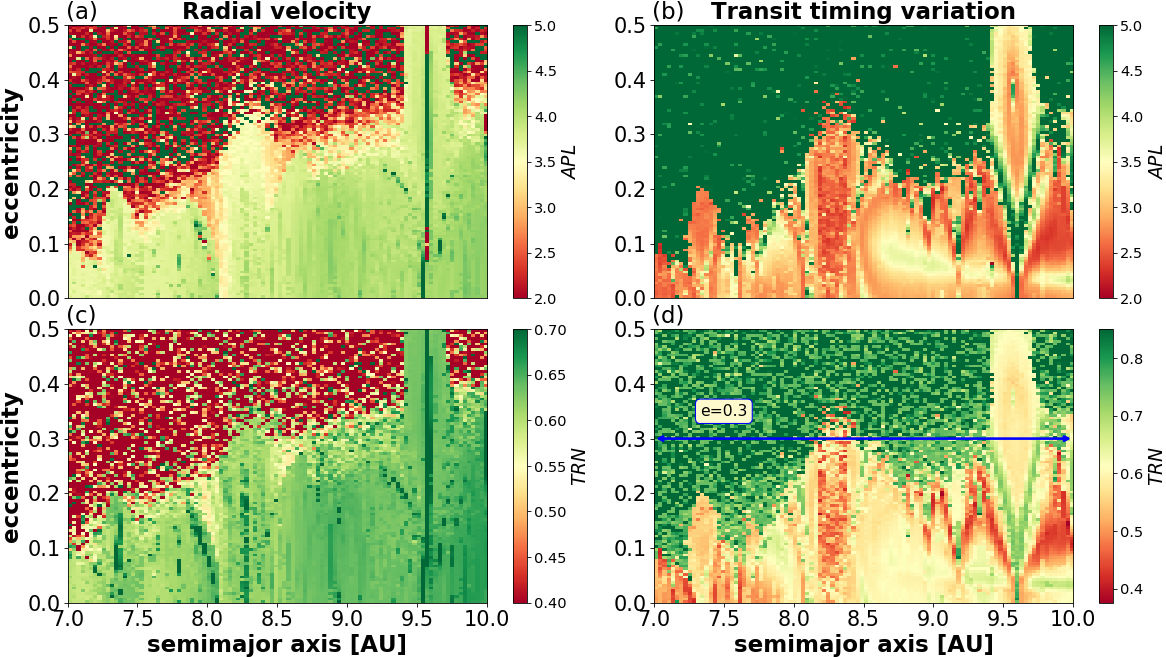}
\caption{Color maps of two RN measures ${APL\text{ and }TRN}$ over the
  initial condition grid ($a_{\text{Saturn}},e_{\text{Saturn}}$). On
  the left the base of the analysis is the Sun's RV while on the right
  it is Jupiter's TTV signal. Every time series contains 950 data
  points. The $e_{\text{Saturn}}$=0.3 line is marked in panel (d) in order to guide the eye in Figure~\ref{fig:rna-noise-gap}. To calculate various RN measures the publicly available \texttt{PYUNICORN} (\url{http://www.pik-potsdam.de/~donges/pyunicorn/}) package has been used \citep{Donges2015}.}  
\label{fig:rna-ae-map}
\end{figure*}
\end{center}

In order to explore different types of dynamical regimes
(quasi-periodic or chaotic) two basic network measures are introduced. Other
fundamental quantitative characteristics of complex networks can be
found in recent groundbreaking works of \citet{Albert2002,Newman2003,Boccaletti2006}. 

The average path length $APL$ can be defined as the arithmetic
mean of geodesic distance ($d_{ij}$) between all pair of vertices
($i,j$) 
\begin{equation}
  APL=\frac{2}{N(N-1)}\sum_{i\neq j}d_{ij},
  \label{eq:apl}
\end{equation}
where $d_{ij}$ is the minimum number of edges between two vertices. In
continuous systems $APL$ is approximately the length of the
trajectory in the phase space, hence, periodic \citep{Zou2010}
orbits are characterized by larger $APL$ than chaotic ones.  

%Before introducing the global clustering coefficient $CLT$ let us describe its local variant $CLT_i.$ The local clustering coefficient gives the probability that two randomly chosen neighbours of a vertex $i$ are also neighbors. That is 
%\begin{eqnarray}
%CLT_{i}&=&\frac{\text{nr. of triangles including vertex
%                   }i}{\text{nr. of triples centered at vertex }i}\\
%  \nonumber 
%&=&\frac{\sum_{j,k}A_{jk}A_{ij}A_{ik}}{\sum_{j,k}A_{ij}A_{ik}(1-\delta_{jk})},
%\label{eq:gcl}
%\end{eqnarray}
%where triple implies a path of length two for which $i$ is the central
%vertex. The global clustering coefficient is the arithmetic average of
%$CLT_i$ over all vertices of the network \citep{Watts1998} 
%\begin{equation}
%CLT=\frac{1}{N}\sum CLT_i.
%\end{equation}
Transitivity ($TRN$), which is a closely related quantity to clustering, is the relative number of triangles compared the total number of connected triples of nodes. In contrast to the global clustering
coefficient, transitivity gives equal weights to all triangles in the network
\citep{Barrat2000}: 
\begin{eqnarray}
TRN&=&\frac{3\times\text{ nr. of triangles in the network
               }}{\text{nr. of linked triples of vertices }}\\
  \nonumber 
&=&\frac{\sum_{i,j,k}A_{jk}A_{ij}A_{ik}}{\sum_{i,j,k}A_{ij}A_{ik}(1-\delta_{jk})}.
\label{eq:trn}
\end{eqnarray}

After the mathematical definitions of RN-related quantities, let us apply them to
the synthetic data sets obtained from the model systems SJS. The
average path length and transitivity has been calculated over the
stability map ($a_{\text{Saturn}},e_{\text{Saturn}}$). Figure~\ref{fig:rna-ae-map} summarizes
the results based on the RV of the Sun and the TTV of Jupiter.

At a glance, it is obvious that all four panels show the structure of
Figure~\ref{fig:megno}. The chaotic and regular parts can be easily
distinguished. Nevertheless, the details around MMRs and at the edge
of chaos also match the texture. Before going into the further details, we should highlight some important results of previous studies. \citet{Zou2010} and \citet{Zou2016} investigated continuous and discrete dynamical systems by means of recurrence network technique. They found the following specific features:
\begin{enumerate}[(i)]
\item Transitivity takes larger value for regular orbits and lower value for chaotic ones both in continuous and in discrete systems.
  \item However, the average path length shows different behavior. For discrete systems, the $APL$ of a periodic orbit is smaller than that of a chaotic one. In contrast, for continuous systems, periodic orbits are characterized by larger average path length than chaotic ones.
\end{enumerate}

%Moreover, \citet{Zou2016} also distinguishes two cases. First, when
%the threshold $\epsilon$ is fixed, second -- that also applies in this
%study -- when the recurrence rate $RR$ is fixed. They claim, based on
%the ''classical'' 2D standard map\footnote{Recurrences in high
%dimensional standard map are presented
%  in Appendix~\ref{app:4Dstm}}, that in these cases the average path
%length behaves differently. More precisely, for a fixed threshold
%$\epsilon$, the average path length depends on the size of the orbit
%in the phase space wherein periodic orbits are confined to
%substantially smaller regions than space filling chaotic
%trajectories. This results in larger values for the latter than that
%of the regular ones. When $RR$ is fixed this difference does not show up.

Now, we can analyse the panels in Figure~\ref{fig:rna-ae-map} in more
details. Let us concentrate on the left column first where the color
maps depict the recurrence network measures $APL$ and $TRN$ based on
the Sun's radial velocity. The results we see here are, loosely speaking, based on a continuous dynamical system. Since the time series are obtained from the RV of the Sun which is a real component of the phase space trajectory. Panel (a) demonstrates that regular motion accomplishes larger APL while chaotic behaviour comes with smaller values. This also applies for the measure transitivity in good agreement with point (i) above.

\begin{center}
\begin{figure*}
  \includegraphics[angle=0,scale=.35]{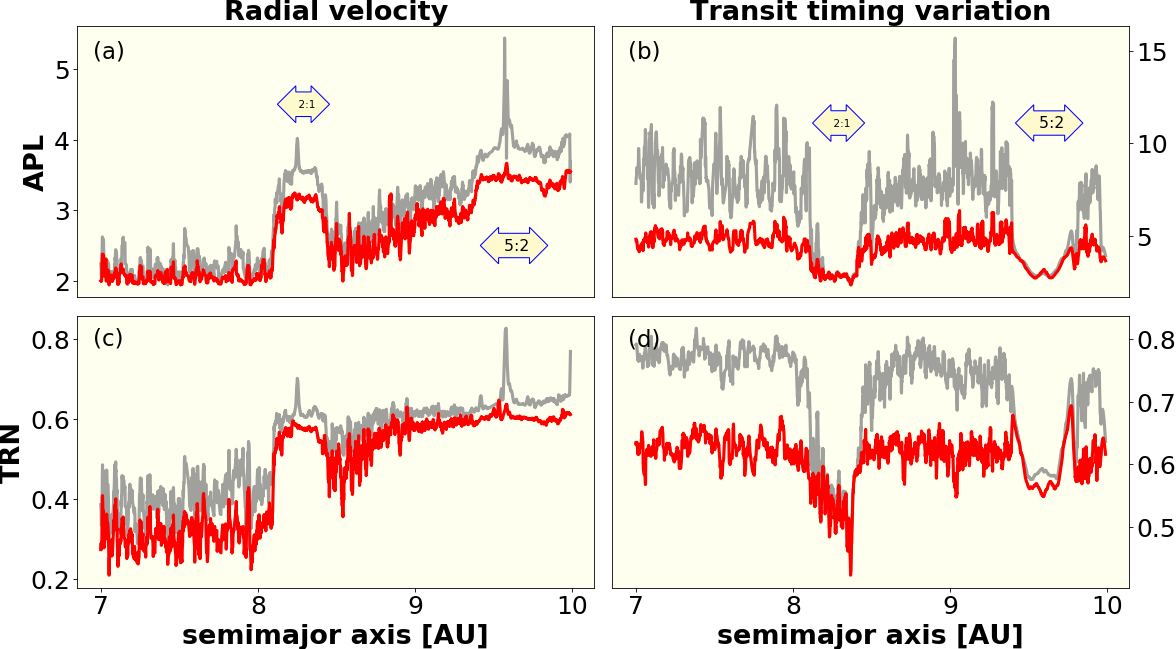}
\caption{RN measures $APL\text{ and }TRN$ versus the initial
  conditions $a_{\text{Saturn}}.$ The other initial condition
  $e_{\text{Saturn}}$ is set to be 0.3 as indicated in Fig.~\ref{fig:rna-ae-map}. The two curves represent the synthetic (gray) and noisy-gappy (red) data sets, respectively.}  
\label{fig:rna-noise-gap}
\end{figure*}
\end{center}

The panels (b,d) belonging to TTVs of Jupiter, i.e. the network
measures acquired from the transit times of the larger planet, can be
interpreted somewhat differently. We should see that transit times
either of Jupiter or Saturn do not represent any component of the
phase space trajectory. They are based on special configurations when
the planet passes in front of the star. This situation can be thought
of as a discrete map\footnote{Well-known maps in dynamical systems
  theory are the stroboscopic and Poincar\'e maps. However, transit
  times do not satisfy the required conditions for these concepts.}
rather than a continuous trajectory, which plays the a crucial role in our following argument. If we consider the TTV signal as a map-like description of a continuous dynamical system, the information in panel (b) is consistent with (ii). Namely for discrete systems $APL$ is smaller for regular motion. Besides, the landscape of the ($a_{\text{Saturn}},e_{\text{Saturn}}$) map shows extremely fine details of the stability regions and MMRs (recall Figure~\ref{fig:megno}). It is clear that measure $APL$ is capable of distinguishing regular and chaotic behavior based on measurable quantities such as RV and TTV in a synthetic two-planet system.

Scrutinizing panel (d) the overall picture is promising since it is
almost identical to panel (b). However, one can observe that the $TRN$
values for regular motion are clearly less than those for chaotic
orbits. This situation completely contradicts point (i) above. To
understand the discrepancy several control computations have been done
including different embedding parameters, longer integration time and
also longer time series with more data points. However, we always find
the same situation, namely regular motion is characterized by lower
transitivity. \citet{Donner2010} and also \citet{Zou2010} underline
that for discrete systems the RN breaks down into disjoint components
around periodic orbits since they appear as finite sets of points in
the phase space. Thus the transitivity tends to 1. This is, however, not the case in higher
dimensions. Because the periodic orbits and trajectories nearby are
not restricted to small domains of the phase space
anymore. Consequently, the regular trajectories can explore regions as
large as the chaotic ones in the phase space. For more details, we
draw the readers attention to Appendix~\ref{app:4Dstm} where we
demonstrate the effect discussed above by using coupled standard maps
as high dimensional discrete Hamiltonian systems. Based on the
previous argument, we point out that the RN-measure transitivity (TRN)
provides smaller value for regular dynamics and larger value for
chaotic motion when analysing the three body problem from discrete
dynamical systems point of view, i.e. when the phase space
reconstruction is based on the TTVs.

\subsection{Noisy time series and missing data points}
\label{sec:noise_gap}

Up to now the analysis has been performed on numerically generated
synthetic time series where the measurements evenly sample the exact
noiseless calculations. However, to demonstrate the robustness of RN
analyses against noise and possible missing data points the
theoretical calculations are modified as follows.

First, a Gaussian white noise, with zero mean and standard deviation
equal to one, is added to the original time series.  The amplitude of
the noise is chosen to be 15\% of the amplitude of the original
signal. Furthermore, in order to imitate astronomical observations
(even those obtained by space based detectors) a certain amount (20\%)
of randomly selected data points have been extracted from the original
time series. More precisely, in the first stage the center of an
interval is randomly defined along the data set, then, in the second
stage, the length of this interval is chosen again randomly. The data
points falling in the interval are removed from the signal. This procedure is repeated until the desired percentage of the missing data is achieved.

In Section~\ref{sec:reconstruction} time delay embedding was
established. This type of reconstruction requires uniformly sampled
time series which is not fulfilled when dealing with scanty data
set. It has been recently shown by
\citet{Lekscha2018} that a cubic spline interpolation of the original data back to uniform time series followed by the classical time delay embedding provides reasonable good phase space reconstructions. According to this, we use in our analysis time delay embedding on cubic splined data sets.

Figure~\ref{fig:rna-noise-gap} illustrates the RN measures $APL$ and
$TRN$ in the noisy SJS model. In order to save computation time only
one section (see the blue arrow in Figure~\ref{fig:rna-ae-map}d) of
the ($a_{\text{Saturn}},e_{\text{Saturn}}$) plane has been
investigated. Each panel contains the results based on the synthetic
time series (gray) as well as the noisy and non-uniformly sampled data
(red). Basically the gray curves correspond exactly to the values in
Figure~\ref{fig:rna-ae-map}. The red dots follow the gray structure
but the contrast is somewhat weaker. That is, the effect of the
presence of noise and missing data points results in smaller
difference between regular and chaotic values of RN
measures. Nevertheless, the ordered motion corresponding to MMRs (2:1
and 5:2) is still perfectly detectable. One can also observe that RN
measures for RV data behave somewhat different than that of TTV,
especially between the two MMRs. And also the same applies here what
has been discussed earlier, namely, lower values of $TRN$ correspond
to regular motion in TTV signals.

We have demonstrated that recurrence network measures are precise and
convenient tools to analyse regular and chaotic motions in
gravitational three body problem. The method is also acceptable in
real world examples when the signal is loaded with noise and the
sampling is not perfectly uniform. Nevertheless, it should be
emphasized that the obtained numbers ($APL,\,TRN$) are \textit{relative 
values} even though they distinguish the different types of motion
correctly. This means, for example, that the average path length for
regular dynamics is smaller than for chaotic (in the case of discrete systems). Therefore, the question arises
naturally how one can decide from one single scalar time series
whether it comes from a chaotic/regular dynamics when we do not have any objective reference point. 
The answer is given by surrogate data analysis described in next section.

\subsection{General principles of surrogate tests}
\label{sec:surrogate}

Surrogate tests are examples of Monte Carlo hypothesis tests \citep{Theiler1992,Theiler1996} applied for testing nonlinearity in a time series. The basic idea is that a nonlinear observable $\lambda_0$ is computed from the data and then one has to decide whether $\lambda_0$ suggests that the data are nonlinear. First, a null hypothesis is taken we want to test, say, the data come from linear processes. Then a number of artificial data sets are generated which are consistent with the null hypothesis. This means they have linear properties similar to those of the original data but nonlinearities are removed. In practice, the newly generated surrogates preserve some properties of the original signal (mean, variance, power spectra). Having the ensemble of surrogate time series discriminating statistics are performed, i.e. the nonlinear observables $\lambda_{i}$ are also computed for the surrogates and then they are compared with the original one. If the value of discriminating statistics $\lambda_0$ from the original time series does not fall within the distribution of the discriminating statistics of the surrogates, the null hypothesis should be rejected. Otherwise, the signal does not contain nonlinearities.

To quantify this process, we can use hypothesis testing \citep{Kantz2003,Schreiber2000} . If we have a good reason to suppose that the distribution of $\lambda_{i}$ is Gaussian, then the mean and standard deviation define the significance which can be used to construct a desired significance level of inference. However, in general, the distribution of discriminating statistics of surrogate data set is not normal, and, therefore, using a rank-based test instead is a better choice. Suppose $N$ surrogate time series are generated and $\lambda_i$ are the calculated nonlinear measures of $i$th ($i$ = 1$\dots,\;N$) data set. Let $\lambda_0$ be the discriminating statistics for the original signal. There are $N$+1 $\lambda$s. Now, all these discriminating statistics are ranked in an increasing order. If the original signal was generated by a linear process, the chance that $\lambda_0$ is the smallest will be 1/($N$+1). According to the rank-order strategy, the null hypothesis is rejected if $\lambda_0$ is the smallest among the ($N$+1)$\lambda$s. This will give a false rejection if $\lambda_0$ being smaller than other $\lambda_i$ by chance, which occurs with probability 1/($N$+1). That is, if we want to have a false rejection with 95\% significance (2 'sigma'), 19 surrogate time series have to be generated in case of one-sided test.

\subsubsection{Pseudo-Periodic Twin Surrogates (PPTS)}
\label{seq:ppts}

Since our method is based on recurrence network analysis, and the exoplanetary observables show mostly quasi-periodic and chaotic behavior, we use, through our analysis, the PPTS method to generate surrogate time series. The PPTS algorithm uses the phase space structure and RP representation to generate surrogates.
%The backbone of the algorithm
%follows the TS method \citep{Thiel2006} , however, for shorter data
%sets, not enough twins can be found. To overcome this disadvantage an
%extra randomization is introduced, Eq.~\ref{eq:noise_radius}. This
%technique comes from the PPS algorithm \citep{Small2001,Luo2005} and maps the phase space points according to their distance. This part is used when no twins are found and generates surrogate that follows approximately the same vector field as the original data but some dynamic noise is added in order to destroy intercycle dynamics \citep{Small2001}. 

In what follows, we give the algorithm of the PPTS \citep{Carrion2016}.
\begin{center}
\mybox{
\begin{enumerate}[(1)]
\item Construct the RP of the original signal using Equation~(\ref{eq:RP}) with predefined $RR$ and identify the twin points. The larger the $RR,$ the more twins. We found the $RR$=0.1-0.15 is adequate in order to find twins.
\item Randomly choose an initial condition $i_0$ and set $i=i_0.$ Initialize $n$=1.
\item If there is a twin point for $\mathbf{x}(i),$ make the next point of the surrogate  $\mathbf{x_{\mathrm{s}}}(n)=\mathbf{x}(j),$ where $j$ is randomly chosen among the twin points with probability 1/$T$ (here $T$ is the number of twins for the state $\mathbf{x}(i)$). Let $i=j$ and $n=n$+1.
\item If there is no twins for $\mathbf{x}(i),$ choose a neighbor $\mathbf{x}(j)$ from all of the elements of the phase space  representation with probability
\begin{equation}
  P(\mathbf{x}(j))\propto \exp{\frac{-||\mathbf{x}(i)-\mathbf{x}(j)||}{\rho}},
  \label{eq:noise_radius}
\end{equation}
where $\rho$ is the noise radius \citep{Small2001}. Set the next point of the surrogate $\mathbf{x_{\mathrm{s}}}(n)=\mathbf{x}(j).$ Let $i=j$ and $n=n$+1.
\item Repeat from (3) and (4) until $n=N.$
\end{enumerate}
}
\end{center}

The crucial step in the PPTS algorithm above is how to choose the
noise radius $\rho$ in Eq.~(\ref{eq:noise_radius}). When $\rho$ is too
small, the original signal and the surrogate are identical, while in
case of too large noise radius the generated surrogates will be
uncorrelated random points. \citet{Small2001} suggest a careful
selection method for $\rho$ when the noise level is fine tuned so that
the intercycle dynamics are destroyed but intracycle dynamics are
preserved. We propose a different method to find the desired
noise radius. The base of our alternative is to measure the similarity
between two time series, called dynamic time warping (DTW). Here, we
only use the outcome, i.e. the value of $\rho,$ of the DTW procedure (a detailed description and example can be found in Appendix~\ref{app:DTW}).

As mentioned above the PPTS method is strongly connected to the
recurrences and the underlying phase space structures. Since recurrence plots determined from PPT surrogate data behave differently for regular and chaotic time series compared to the RPs obtained from the original signal, one can test \textit{the null hypothesis that the observed time series is consistent with quasi-periodic orbit.}

\begin{center}
\begin{figure}
  \includegraphics[angle=0,width=\columnwidth]{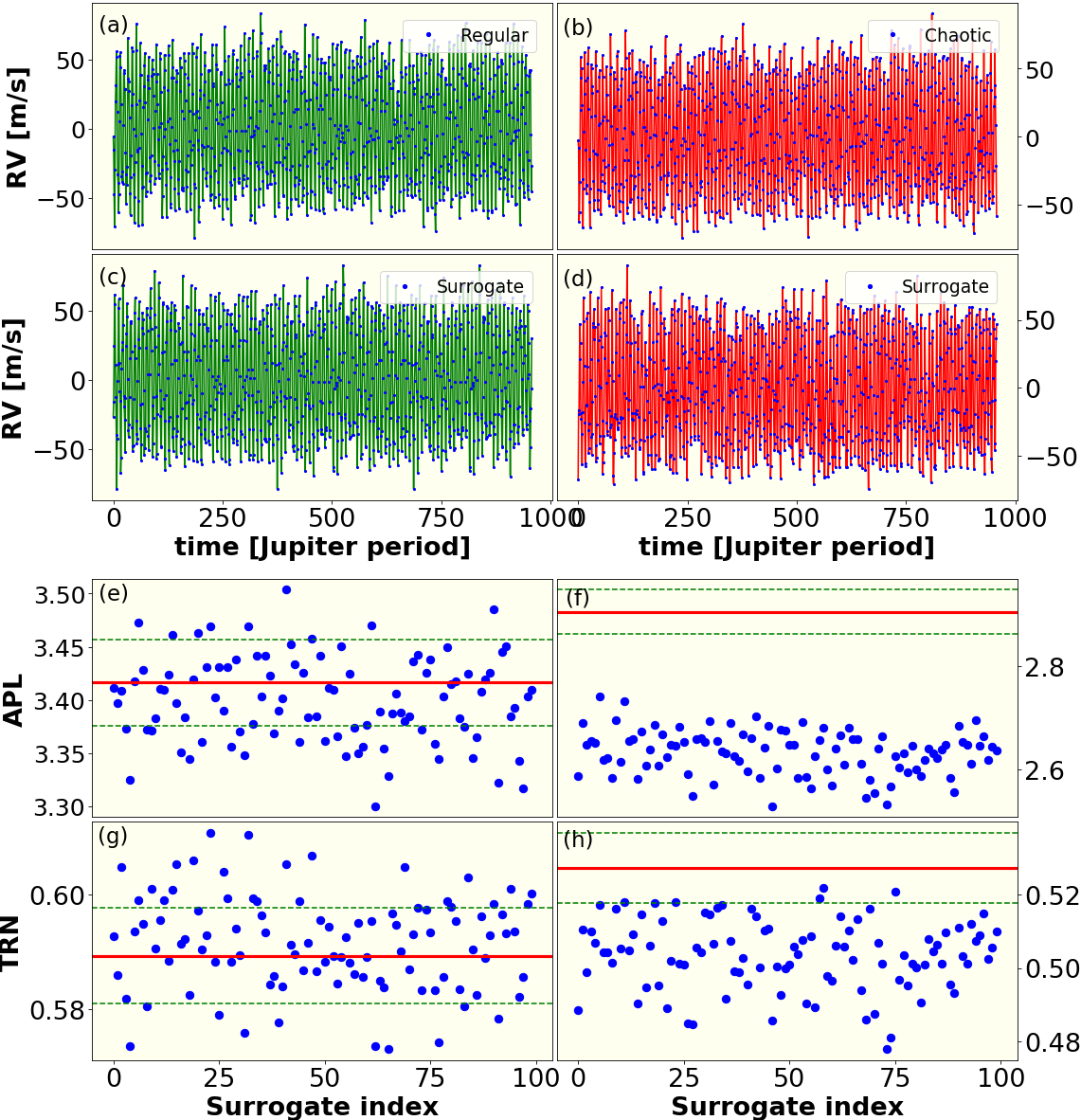}
\caption{Hypothesis test based on the RV signal in the SJS
  system. Left: Regular dynamics - original time series (a), one of
  the surrogate data sets (c), different recurrence network measures
  ($APL,$ $TRN$) compared to the same characteristics calculated from
  PPTSs. The noise radius appearing in the PPTS algorithm is $\rho_{\mathrm{stable}}=0.202.$ Right: Chaotic dynamics - panels match those on the left. The green dashed lines denote the $\pm$1 standard deviation of blue data points around the red line (RN measures of the original signals). $\rho_{\mathrm{chaotic}}=0.252.$}  
\label{fig:conf-rv}
\end{figure}
\end{center}

\begin{center}
\begin{figure}
  \includegraphics[angle=0,width=\columnwidth]{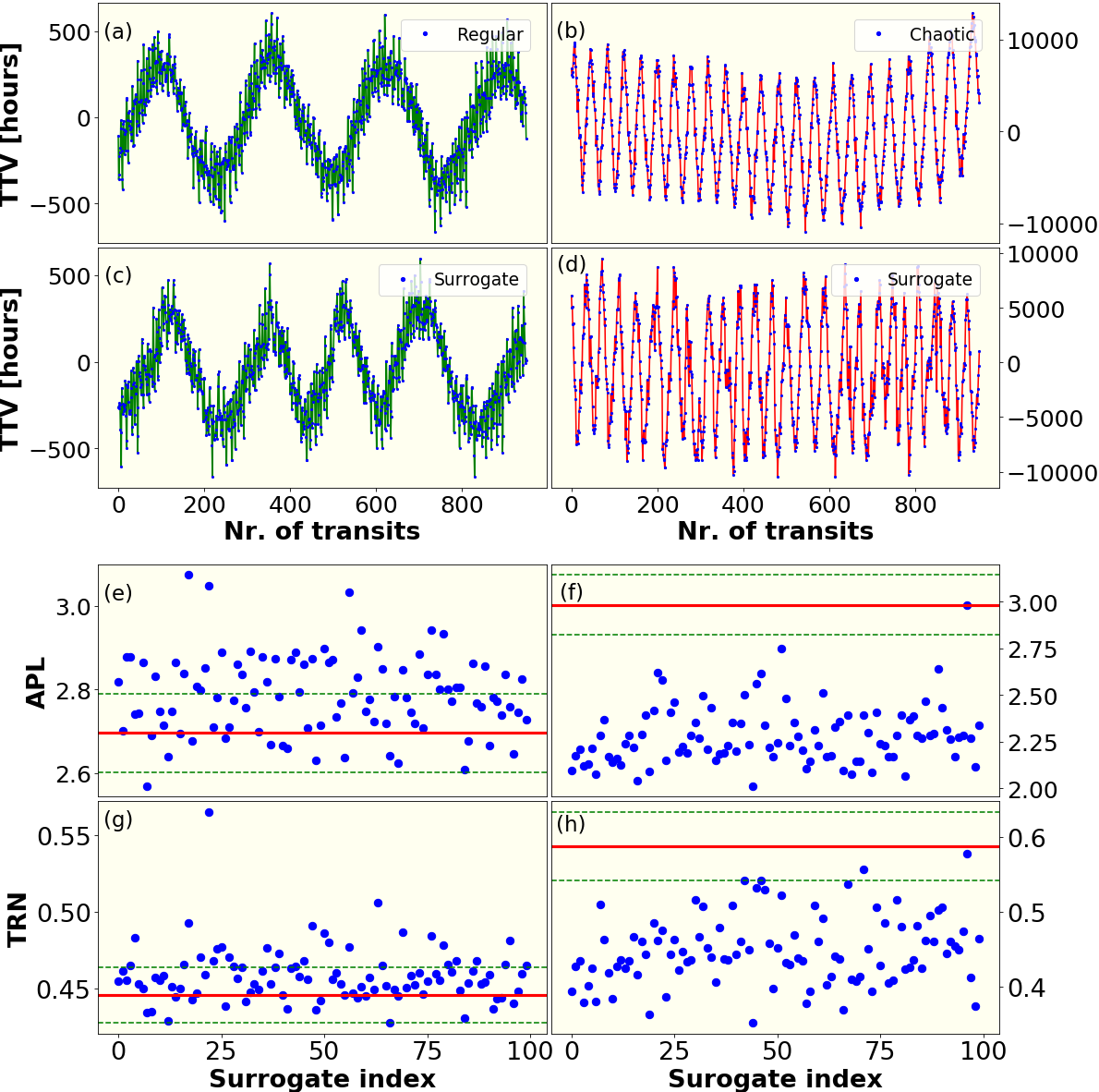}
  \caption{Hypothesis test exploration of (Jupiter) TTV signal. The panels have the content as in Figure~\ref{fig:conf-rv}. $\rho_{\mathrm{stable}}=0.089$ and $\rho_{\mathrm{chaotic}}=0.13,$ see Appendix~\ref{app:DTW}.}
\label{fig:conf-ttv}
\end{figure}
\end{center}

Now, let us consider several examples from the model system SJS to see how PPTS works in practice. We pick up two initial conditions from the ($a_{\text{Saturn}},e_{\text{Saturn}}$) parameter plane one regular and one chaotic as usual. First, we deal with RV data. Figure~\ref{fig:conf-rv}(a) and (b) portray the RV time series, i.e. Sun's velocity component $v_x,$ for initial conditions $(a_{\text{Saturn}},e_{\text{Saturn}})=(7.2,0.02)$ and $(a_{\text{Saturn}},e_{\text{Saturn}})=(8.0,0.2),$ similarly as in Fig.~\ref{fig:RV_TTV}. These two initial conditions are also marked as blue triangles in Figure~\ref{fig:megno}. Both data sets contain 950 points, the difference to those in Figure~\ref{fig:RV_TTV} is that current time series are cubic splined (due to missing observations) with additional noise. According to the description above we generated 100 surrogates for regular as well as chaotic time series, respectively in order to achieve 99\% significance in hypothesis test. Panels (c) and (d) illustrate one pair of the corresponding pseudo-periodic twin surrogates.

Having the surrogate data sets we can compare the RN measure of the
original signal and those obtained from PPTSs. We focus on $APL$ and $TRN$
as before. The bottom four panels, Figure~\ref{fig:conf-rv}(e)-(h),
show the relation of these values. In each plot the red solid line
represents the $APL$ and $TRN$ values of the original signal. The blue
circles correspond to the $APL$s and $TRN$s for 100 different PPTS data
sets. The rank-based test reveals that in the regular case both $APL$
and $TRN$ fall into the zoo of surrogate RN measures. In contrast, when
the dynamics is chaotic the original measures are located well outside
the set of surrogate points. That is, in the later case we can reject
the null hypothesis, according to which the original signal comes from quasi-periodic motion.

One can accomplish the same analysis on TTV data sets as
well. Figure~\ref{fig:conf-ttv} depicts the hypothesis test for the
same two orbits based the TTV signal of Jupiter\footnote{The method
  works well for Saturn's TTV too.}. We call again the readers'
attention that 950 transits cover slightly different time spans for different
types of motion. 

\begin{center}
\begin{figure}
  \includegraphics[angle=0,width=\columnwidth]{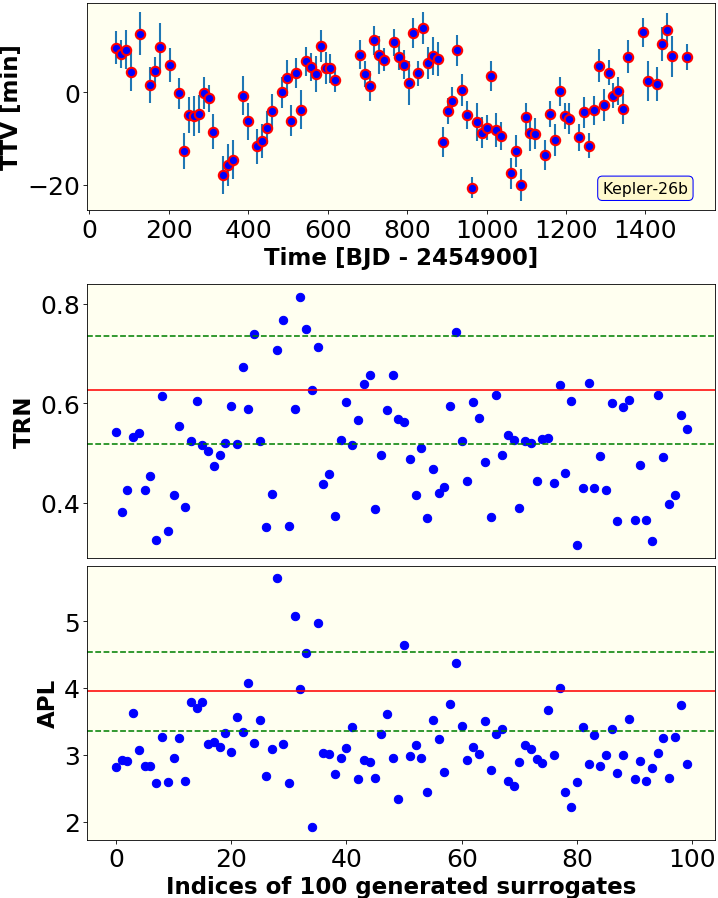}
  \caption{TTV signal (top) and hypothesis test (middle and bottom) of Kepler-26b. The horizontal solid line represents the $TRN=0.627$ and $APL=3.951$ for the original time series, while the dashed lines show the $\pm$1 standard deviation ($\rho=2.076$) of the same measures calculated from 100 PPTSs (circles). }
\label{fig:kepler-26b}
\end{figure}
\end{center}

%\begin{center}
%\begin{figure}
%  \includegraphics[angle=0,scale=.35]{./noise_radius_koi-250_01_lowres.png}
%  \caption{Hypothesis test exploration of (Jupiter) TTV signal. The panels have the content as in Figure~\ref{fig:conf-rv}.}
%\label{fig:conf-ttv}
%\end{figure}
%\end{center}

%\begin{center}
%\begin{figure}
%  \includegraphics[angle=0,width=\columnwidth]{./Confidence_PPTS_koi-250_01_lowres.png}
%  \caption{Hypothesis test of Kepler-26b TTV signal. The horizontal solid line represents the $TRN=0.627$ (top) and $APL=3.951$ (bottom) for the original time series, while the dashed lines show the $\pm$1 standard deviation ($\sigma=2.076$) of the same measures calculated from the PPTSs (circles). }
%\label{fig:kepler-26b-test}
%\end{figure}
%\end{center}

\begin{center}
\begin{figure}
  \includegraphics[angle=0,width=\columnwidth]{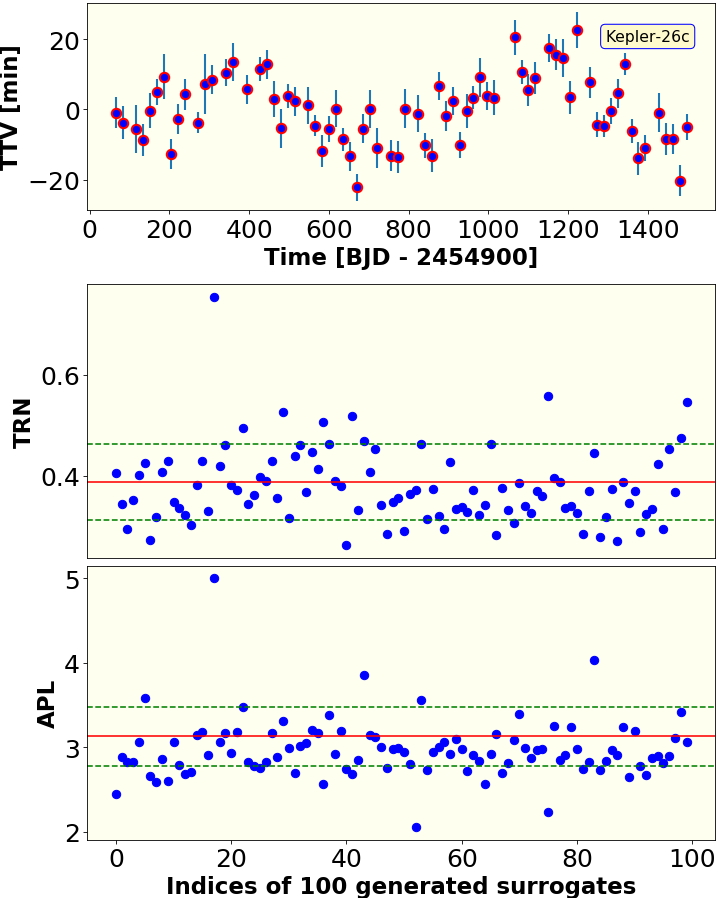}
  \caption{Top: Kepler-26c TTV. Middle-bottom: Hypothesis test for Kepler-26c. $TRN=0.387$ , $APL=3.133$ , $\rho=2.977$.}
\label{fig:kepler-26c}
\end{figure}
\end{center}

%\begin{center}
%\begin{figure}
%  \includegraphics[angle=0,scale=.35]{./noise_radius_koi-250_02_lowres.png}
%  \caption{Hypothesis test exploration of (Jupiter) TTV signal. The panels have the content as in Figure~\ref{fig:conf-rv}.}
%\label{fig:conf-ttv}
%\end{figure}
%\end{center}

%\begin{center}
%\begin{figure}
%  \includegraphics[angle=0,width=\columnwidth]{./Confidence_PPTS_koi-250_02_lowres.png}
%  \caption{Hypothesis test for Kepler-26c. $TRN=0.387$ , $APL=3.133$ , $\sigma=2.977$.}
%\label{fig:kepler-26c-test}
%\end{figure}
%\end{center}

\section{Application to exoplanetary systems}
\label{sec:data}

The results presented in previous sections are based on numerical
integration of a well-defined planetary system including the Sun,
Jupiter, and Saturn. Our RN analysis shows that 950 data points either
RV or TTV measurements (in case of Jupiter) are enough to carry out
the stability investigation of the system. Next we want to apply the
whole procedure to real exoplanetary systems that are known in the literature. As always, the amount and quality of the acquired data is extremely important, we try to analyse the best public data sets. Thus, we decide to use only space-based TTV signals, e.g. Kepler data, since the available RV measurements about two-planet systems contain a small number of data points and are very sparse in time.

\subsection{Data}

After 17 quarters the \textit{Kepler satellite} finished its original
mission and collected more than 69,000 transits for 779 KOIs with high
signal to noise ratio (SNR). The most interesting systems with
significant long-term TTVs have been pilled up and published in a
catalog (\url{ftp://wise-ftp.tau.ac.il/pub/tauttv/TTV/ver\_112})
\citep{Holczer2016} in order to make the light curves more usable for
further research. We limited the choice of possible systems to those
presented in \citet{Holczer2016} and their stability analysis can be
found in a recent paper of \citet{Panichi2018} for comparison.

\subsubsection{Four-planet system: Kepler-26} \label{sec:kepler26}

The stability analysis of this planetary system \citep{Steffen2012} is
based on the TTV signals of two super-Earths (out of the confirmed
four by \citet{Jontof2016}) being in 7:5 MMR. First, we have done the
cubic spline fit since the time series has 17~\% missing transits in
the available epoch frame. Then after the phase space reconstruction
the noise radius has been obtained as, $\rho=2.076.$ Having the value of $\rho$ we can generate the surrogates and perform the hypothesis test. Figure~\ref{fig:kepler-26b} shows the TTV signal and the results of the hypothesis test for the inner planet, Kepler-26b. The two bottom panels portray transitivity ($TRN$) and average path length ($APL$) of recurrence network (red solid lines). The same measures of the 100 surrogates (blue circles) encompass those coming from the original signal, the null hypothesis can be accepted, i.e. the observed time series is produced by quasi-periodic dynamics.

The same applies to Kepler-26c. The missing transit events give a
signal of 14~\% of the whole covered time span. Based on the
hypothesis test one can conclude that the planetary dynamics shows
regular motion, see Figure~\ref{fig:kepler-26c}. 

\begin{center}
\begin{figure}
  \includegraphics[angle=0,width=\columnwidth]{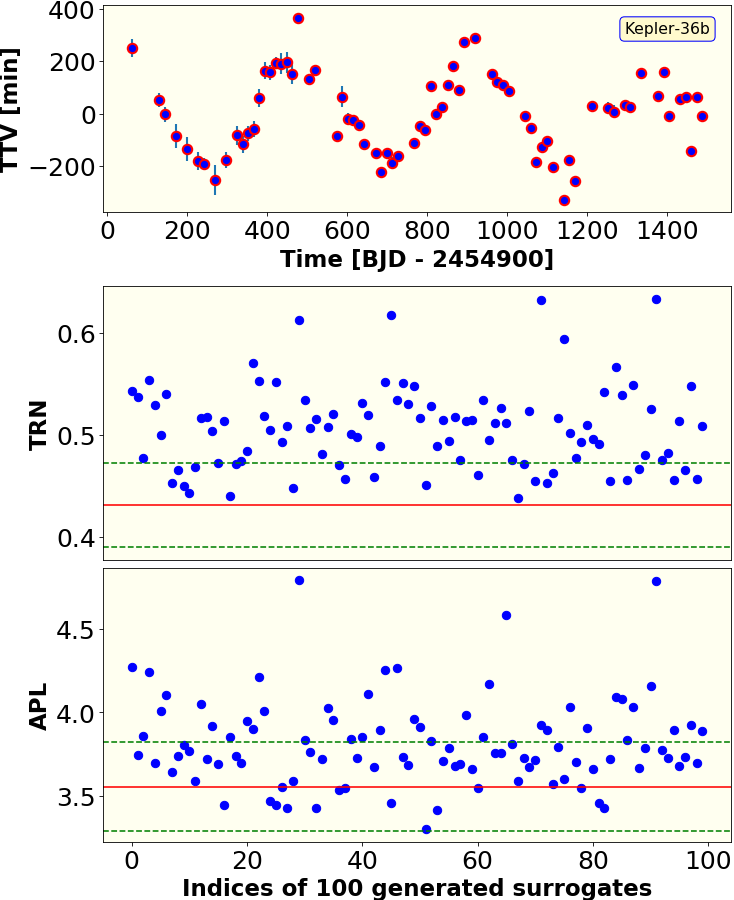}
  \caption{TTV of Kepler-36b  (missing data 13~\%) and hypothesis test. The panels have the content as in Figure~\ref{fig:conf-rv}. $TRN=0.430$ , $APL=3.557$ , $\rho=37.177.$ Credit: \citet{Kovacs2019}.}
\label{fig:kepler-36b}
\end{figure}
\end{center}

%\begin{center}
%\begin{figure}
%  \includegraphics[angle=0,scale=.35]{./noise_radius_koi-277_01_lowres.png}
%  \caption{Hypothesis test exploration of (Jupiter) TTV signal. The panels have the content as in Figure~\ref{fig:conf-rv}.}
%\label{fig:conf-ttv}
%\end{figure}
%\end{center}

%\begin{center}
%\begin{figure}
%  \includegraphics[angle=0,width=\columnwidth]{./Confidence_PPTS_koi-277_01_lowres.png}
%  \caption{Hypothesis test exploration of (Jupiter) TTV signal. The panels have the content as in Figure~\ref{fig:conf-rv}. $TRN=0.546$ , $APL=5.75$ , $\sigma=20.381$.}
%\label{fig:kepler-36b-test}
%\end{figure}
%\end{center}

\begin{center}
\begin{figure}
  \includegraphics[angle=0,width=\columnwidth]{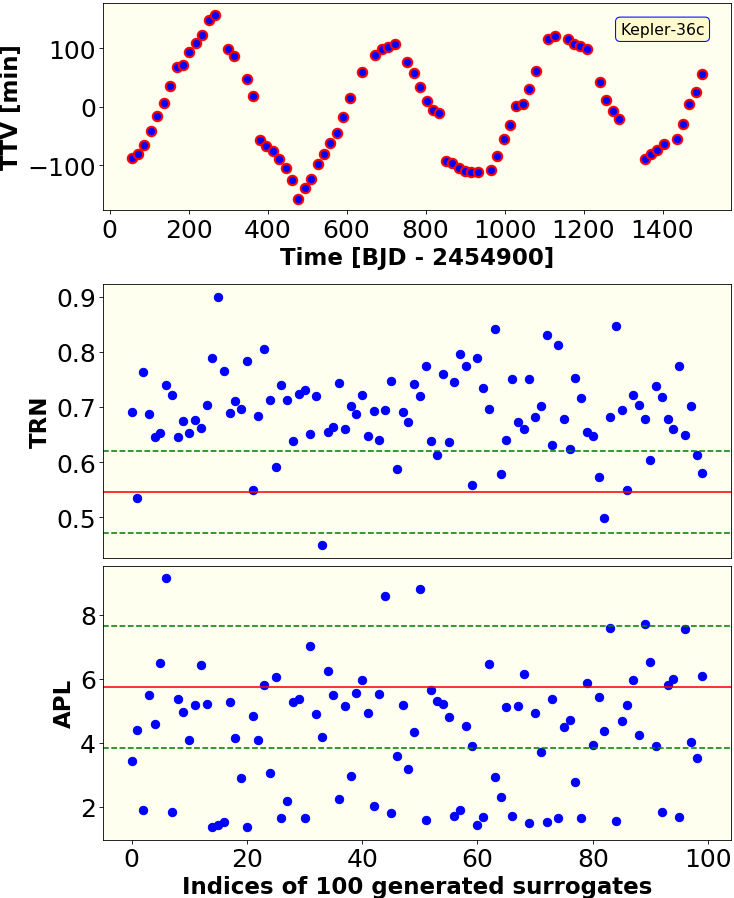}
  \caption{TTV of Kepler-36c (missing data 30~\%). Hypothesis test. $TRN=0.546$ , $APL=5.75$ , $\rho=20.381$. Credit: \citet{Kovacs2019}.}
\label{fig:kepler-36c}
\end{figure}
\end{center}

%\begin{center}
%\begin{figure}
%  \includegraphics[angle=0,scale=.35]{./noise_radius_koi-277_02_lowres.png}
%  \caption{Hypothesis test exploration of (Jupiter) TTV signal. The panels have the content as in Figure~\ref{fig:conf-rv}.}
%\label{fig:conf-ttv}
%\end{figure}
%\end{center}

%\begin{center}
%\begin{figure}
%  \includegraphics[angle=0,width=\columnwidth]{./Confidence_PPTS_koi-277_02_lowres.png}
%  \caption{Hypothesis test exploration of (Jupiter) TTV signal. The panels have the content as in Figure~\ref{fig:conf-rv}.$TRN=0.430$ , $APL=3.557$ , $\sigma=37.177$}
%\label{fig:kepler-36c-test}
%\end{figure}
%\end{center}

\subsubsection{Two-planet system: Kepler-36} \label{sec:kepler36}

The Kepler-36 system \citep{Deck2012} has one of the largest TTV
signals among the known planetary configurations. Based on this fact,
the dynamical analysis, derived from TTVs of two planets, assisted to
explore a complex behaviour in this extrasolar planetary system. A
great success of the exploration of Kepler-36 is the emergence of
stable chaos.\footnote{A phenomenon related to small range of
  oscillations in proper orbital elements while the Lyapunov time is
  much shorter.} In our analysis the hypothesis test shows no consistent
results. Neither for the two component of the system nor for the
different network measures. As one can see, the original measure of
$TRN$ appears to be the lowest one in the rank based test middle panel
of Figure~\ref{fig:kepler-36b}. This means that the null-hypothesis
can be rejected, i.e. the original signal comes with 99\% level from
chaotic dynamics rather than quasi-periodic. The other measure, $APL$,
shows the opposite (bottom panel) yielding that we can expect a
regular motion. This is, however, not
the case for Kepler-36c (Figure~\ref{fig:kepler-36c}). Kepler-36c appears to be stable for
both $TRN$ and $APL$ in hypothesis test. In fact, most of the surrogate
transitivity values are above the $TRN$ of the original time series
(middle panel of Fig.~\ref{fig:kepler-36c}). Still following the
previous rules we accept the null-hypothesis. Although, three out of four tests characterize the
system as a regular one, the discrepancy definitely shows more
uncertainty. What we can conclude based on the SJS model system and
its hypothesis test, see Figure~\ref{fig:conf-ttv} is that the motion
taking place in Kepler-36 is very close to the resonance border just
like the chaotic ($a_{\mathrm{Saturn}},e_{\mathrm{Saturn}}$)=(8.0,0.2)
pair in SJS system. The stable chaos scenario suggested by
\citet{Deck2012} completely overlaps with the stickiness effect
appearing at the border of regular domains (MMRs in celestial
mechanics) in dynamical systems \citep{Tsiganis2000}. In addition,
\citet{Panichi2018} also found that the system (Kepler-36) is very
close to the border of the 7:6 MMR.

\subsubsection{Two-planet system: Kepler-29} \label{sec:kepler29}

\begin{center}
\begin{figure}
  \includegraphics[angle=0,width=\columnwidth]{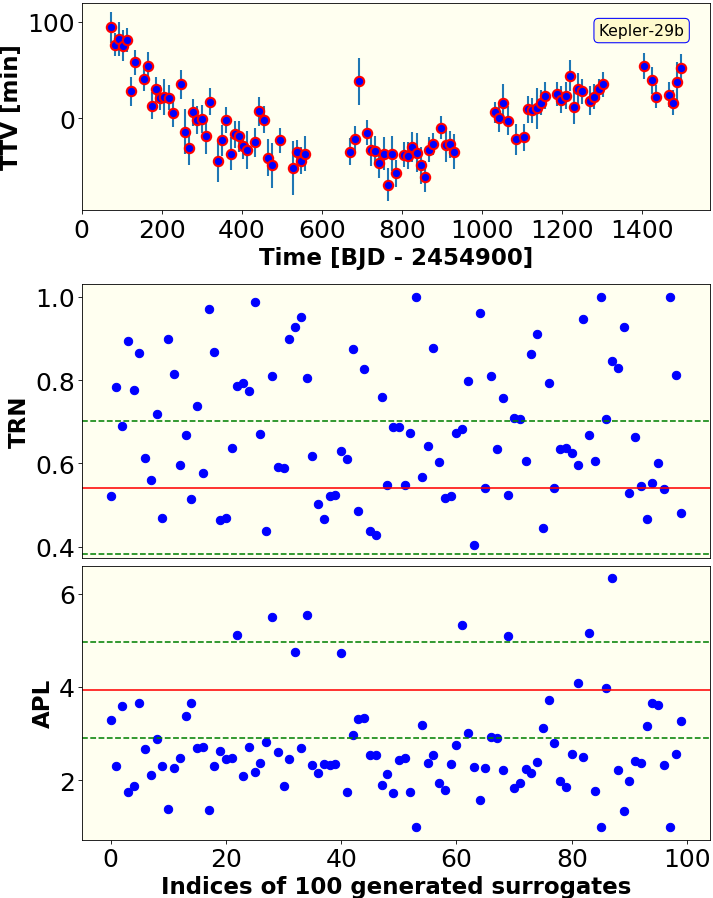}
  \caption{TTV Kepler-29b (missing data 32~\%). Hypothesis test. $TRN=0.541$ , $APL=3.932$ , $\rho=12.6.$}
\label{fig:kepler-29b}
\end{figure}
\end{center}

\begin{center}
\begin{figure}
  \includegraphics[angle=0,width=\columnwidth]{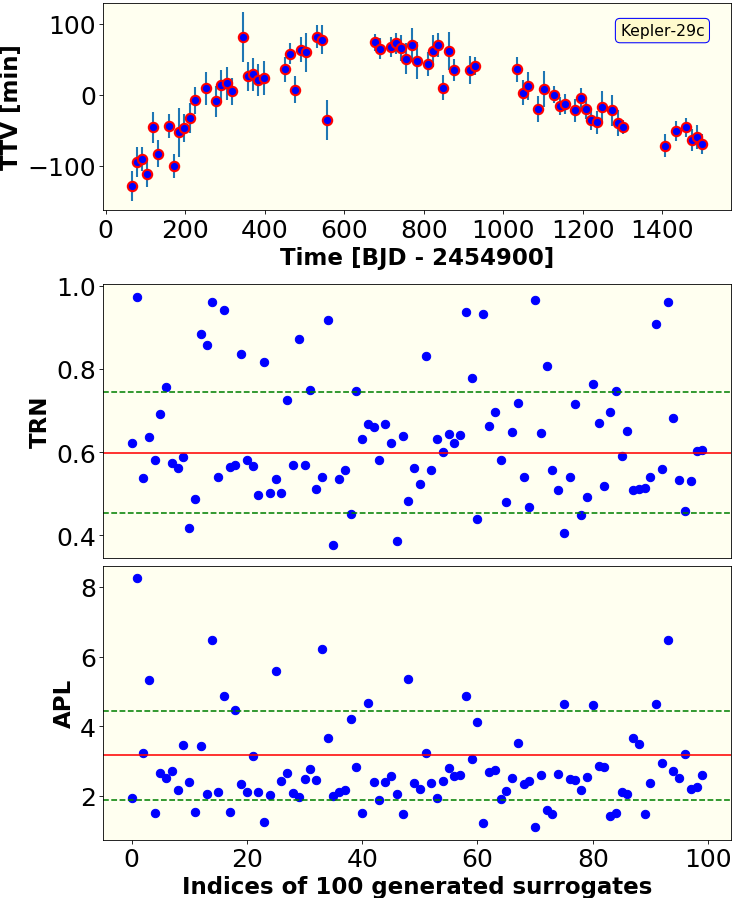}
  \caption{TTV Kepler-29c (missing data 36~\%) and hypothesis test. $TRN=0.598$ , $APL=3.161$ , $\rho=14.21.$}
\label{fig:kepler-29c}
\end{figure}
\end{center}

The Kepler-29 system \citep{Fabrycky2012} harbours two super-Earth
planets on tightly separated resonant orbits. This indicates a fairly
large TTV signal plotted in Figs.~\ref{fig:kepler-29b} and
\ref{fig:kepler-29c} upper panels. Unfortunately, the data is very
sparse compared to other time series in the catalog. The 1500 days run
contains only 93 valuable transit measurements (from possible 108)
producing a signal with 32-36\% missing points. Unfortunately, the
large gaps that can cause trouble and misinterpretation of cubic
splined time series. Thus one should handle the analysis with
care. \citep{Panichi2018} showed that the integration time plays an
important role in the stability investigation of this particular systems. Although, the initial conditions of the inner planet (Kepler-29b) are sitting well inside the regular domain of the ($a,e$) stability map for shorter time, it turns out that the center of the resonance becomes unstable for longer integration. The initial conditions eventually appear at the border of the regular and chaotic domain falling into the sticky region, similarly to Kepler-36.

The results of the hypothesis tests based on the available data sets are shown in Figures~\ref{fig:kepler-29b} and \ref{fig:kepler-29c} predict regular dynamics for the inner as well as for the outer planet. This is not in contradiction to previous findings bearing in mind that the time series is extremely short. It might present chaotic nature for longer time scales. Moreover, in case of Kepler-36 system the planets are closer to each other with an order of magnitude resulting in much stronger mutual gravitational perturbation and dynamical effect on shorter times. Thus, the different dynamical time scales can explain the more irregular outcome of the stability investigation of Kepler-36 from the same kind (length, noise, gaps) of data set.

%\begin{center}
%\begin{figure}
%  \includegraphics[angle=0,scale=.35]{./noise_radius_koi-738_01_lowres.png}
%  \caption{Hypothesis test exploration of (Jupiter) TTV signal. The panels have the content as in Figure~\ref{fig:conf-rv}.}
%\label{fig:conf-ttv}
%\end{figure}
%\end{center}

%\begin{center}
%\begin{figure}
%  \includegraphics[angle=0,width=\columnwidth]{./Confidence_PPTS_koi-738_01_lowres.png}
%  \caption{Hypothesis test exploration of (Jupiter) TTV signal. The panels have the content as in Figure~\ref{fig:conf-rv}.$TRN=0.541$ , $APL=3.932$ , $\sigma=12.6$}
%\label{fig:kepler-29b-test}
%\end{figure}
%\end{center}

\section{Conclusions}
\label{sec:conclusion}

In this paper we describe the stability of multiplanetary systems
based solely on astronomical observations. The technique involves
complex network analysis through phase space reconstruction (time
delay embedding) and recurrence plot representation (Figure~\ref{fig:work-flow}). The Sun-Jupiter-Saturn system is used for
pedagogical purposes. This well-known two-planet configuration serves
a large variety of motions (periodic, quasi-periodic, chaotic, sticky)
what can also be observed in extrasolar planetary systems. Having the
knowledge about the SJS dynamics (indicators MEGNO) we are able to test
the new strategy to obtain the stability and explore the strength,
weakness, and applicability of the method. We have shown that complex
network measures (such as average path length, transitivity) applied in
recurrence network (RN) framework characterize the system dynamics
accurately. Moreover, our strategy copes with noisy and non uniformly
sampled time series as well. The whole procedure has been applied in
real exoplanetary systems. In spite of short and sparse data sets the
analyses match with earlier stability studies of the same systems. 

Although the long-term stability is the primary goal of dynamical
analysis it should be noted that the method we present is strictly
based on the signal measured. It means no further temporal extension
of the phase space trajectory is available, say for millions of orbital revolutions, like in the case of numerical
integration. Numerical integration is , however, the conventional approach that uses the
orbital parameters as initial conditions obtained from the best
fitting planetary models. And then applying one of the chaos indicator
methods to find long time behaviour of the system. Our technique describes,
in contrast, the dynamical behaviour of reconstructed phase space
trajectory based on the time frame of the available
observations. Clearly, for the direct problem (i.e. numerical
integration of the initial conditions), one can find different
stability phenomenon for slightly modified initial conditions or for
the same initial states but different evolution time. For example,
high eccentricity motion often leads to escape from the system that
occurs for sooner or later. Obviously, no chaos detecting method can
obey such a situation properly. Furthermore, close to the border of
resonances, where stickiness acts, the orbit might exhibit regular
situation, however, after a time it detaches from that domain and
behaves chaotically (or escapes). In this study we show how the method
works for a certain time interval (1050 orbits for Jupiter sampled by 950 measured points) in case of
RV and the same number of transits (950). We note that the method was tested for longer and shorter integration times. That is, the MEGNO map for the same (a,e) parameter plane has been calculated for different time intervals and, consequently, its structure was different, especially at MMRs. Nevertheless, NR measures were also capable in these situation to catch the distinct dynamical situations. 

Let us draw a parallel between the direct numerical integration and RN
method. The best fit planetary model permits accurate initial
parameters for further numerical integration. However, these initial
conditions strongly depend on the length of the time series as well as
S/N. Fine tuning the initial parameters by improving these
circumstances (follow-up observations) might lead to different kind of dynamics than those
emerging from the ''obsolete'' orbital elements. Exactly the same
situation holds for RN measures. The more longer the data set,
i.e. covers larger segments of the dynamics, the precise the stability analysis. Indeed, no extrapolation for the far future is needed and also the whole procedure does not require neither the initial conditions nor the system parameters and the equations of motion explicitly. All in all, what we can do with MEGNO for a given integration time, RN measures do the same job for identical length of time series as well.

To our best knowledge, this study is the first step of application of nonlinear time series analysis methods to obtain stability of exoplanetary systems. Hence, there are many ways to improve it. We mention some of them based on our experience so far.

\begin{itemize}
\item Length of the time series: Long-term stability analysis requires sufficiently long phase space trajectories to determine either the LCEs or other more efficient\footnote{More efficient means in this sense that the method can distinguish chaotic and regular motion from short (few thousands of orbital period) integration times.} chaos detection quantities. This is also true for indirect methods as well. Fortunately, we can expect that the signals will be more populated and precise in the future.
\item Combining RV-TTV-ASTROMETRY data: Various system information acquired by several observation techniques can describe distinct physical actions. The above mentioned observables give us different insights into the dynamics. Obviously, these are coupled since they come from the same system. The recurrence network method used in this work allows to mix different kind of measurements\footnote{This is true only when some basic criteria about the time series are fulfilled. See more about joint recurrence plots and cross-recurrence plots. \citep{Marwan2007}} and with the help of this more information about the system dynamics can be obtained.
\item Different surrogate algorithms: Surrogation is a widely used technique in time series analysis. In this study we performed the analysis by using the PPTS algorithm. However, there are other methods that are probably more sensitive, say, for the linear colored observational noise added to the system \citep{Luo2005}. Therefore, it is worth checking different surrogate generation methods to distinguish chaotic and regular motion in dynamical systems. Furthermore, we used only one static significance test (based on 100 surrogates which yields 99\% level). One also might carry out more tests. In these situations the total probability of a false rejection can be different. To deal with such a scenario and how to modify the significance level we refer to the textbook \citep{Kantz2003}.
\item Different embedding techniques: Phase space reconstruction is a
  must in RN analysis. Besides the classical time delay embedding
  there are various techniques that guarantee a successful trajectory
  reconstruction \cite{Lekscha2018}. Moreover, one can also find alternative parameter estimation of delay embedding \citep{Small2004,Hirata2006,Nichkawde2013} as well as different kind of reconstruction methods \citep{Hirata2017,Uzal2011,Carroll2018a,Lekscha2018} that can also be tested in dynamical astronomy.
\item Other types of networks: Once we have the reconstructed phase space trajectory the natural choice is to use RN measures. However, other types of networks are suitable to explore the dynamical variability. Visibility graphs \citep{Lacasa2008,Zou2014,Mutua2016} are promising candidates to use them as an alternative approach in planetary dynamics.
\item Machine learning techniques: Many papers appeared
  recently about the attractor reconstruction and exploration of chaos
  in dynamical systems making use of machine learning techniques
  called reservoir computing
  \citep{Pathak2017,Lu2018,Nakai2018,Carroll2018b}. In this process
  the input data are the measured time series like in our present
  analysis. Moreover, the method could also solve the problem of
  surrogate analysis which is naturally encoded (as modified
  autonomous reservoir) in the mechanism they use. An additional
  method that might fit well to our purposes is convolutional networks
  of 2D image processing. A fresh paper \citep{Hatami2018} proposed a
  strategy wherein a convolutional neural network (CNN) classifier has
  been applied to recurrence plots obtained from time series of
  dynamical systems. They assert that CNN model works better for
  texture images (practically RPs) than other time series
  classification schemes. These methods hold a significant potential
  to improve future dynamical modeling not only for planetary
  sciences but other fields of physics.
  \item High dimensional Hamiltonian problems: Not only for recurrence
    based analysis but for other network measures most of the analysis
    covers the well-known didactic examples such as dissipative
    R\"ossler, Lorenz, H\`enon-Heiles systems or the classical
    Standard map as a Hamiltonian example. It would be extremely beneficial to examine how the network measures behave in more complex phase space e.g. in high dimensional Hamiltonian systems.
%- to gain other information than stability - system parameters (Kovacs)
\end{itemize}
%=============================================

In summary, we believe that the method presented above can be used as a completion or prerequisite of dynamical analysis based on best fit planetary models followed by numerical N-body integration. Furthermore, recent efforts \citep{Carter2013,Deck2014,Dajka2018} show significant improvement in indirect stability and dynamical analysis based on measured time series that also supports our strategy. 

\section*{Acknowledgements}

The author wish to thank John Chambers his valuble comments and suggestions that result in a more coherent version of the manuscript. This work was supported by the NKFIH Hungarian Grants K119993, PD121223. The support of Bolyai Research Fellowship and UNKP-19-4 New National Excellence Program of Ministry for Innovation and Technology is also acknowledged.

%%%%%%%%%%%%%%%%%%%%%%%%%%%%%%%%%%%%%%%%%%%%%%%%%%

%%%%%%%%%%%%%%%%%%%% REFERENCES %%%%%%%%%%%%%%%%%%

% The best way to enter references is to use BibTeX:

\bibliographystyle{mnras}
\bibliography{RecNet} % if your bibtex file is called example.bib

%%%%%%%%%%%%%%%%%%%%%%%%%%%%%%%%%%%%%%%%%%%%%%%%%%

%%%%%%%%%%%%%%%%% APPENDICES %%%%%%%%%%%%%%%%%%%%%

\appendix

\section{Time delay embedding in practice}
\label{app:delay_example}

Let us consider a heuristic example for the delay reconstruction using
synthetic measurements.% in SJS system, $(a_{\text{Saturn}}, e_{\text{Saturn}})=(8.7,0.3)$.
The univariate data set, containing 600 points, is given as 

\begin{equation}
x(t_i)=\left(
\begin{array}{c}
\#1\\\#2\\\#3\\\vdots\\\#598\\\#599\\\#600
\end{array}
\right) ,
 \quad i=1\dots 600. \nonumber
\end{equation}

Next, choose the embedding parameters, for instance, $m=3$, $\tau=4.$ Consequently,
the size of the matrix $\mathbf{x}_N$ is $3\times 592$ as it can be seen
below

\begin{equation}
\mathbf{x}_N =\{x(t_i-2\cdot 4) ,x(t_i-4) ,x(t_i)\}
= \left(
\begin{array}{ccc}
\tikzmarkin{a}(0.1,-0.1)(-0.1,0.35)\#9& \#5&\#1\tikzmarkend{a}\\
\#10 &  \#6 &\#2\\
\#11&\#7&\#3\\
\vdots&\vdots&\vdots\\
\#598&\#594&\#590\\
\#599&\#595&\#591\\
\#600&\#596&\#592\\\hline
&\vdots&\vdots\\
&\#600&\vdots\\
&&\#600
\end{array}     \right)  . \nonumber \\
%https://tex.stackexchange.com/questions/40028/highlight-elements-in-the-matrix
\end{equation}

Those elements that remain under the horizontal line are omitted.
In this method the recovered signals are in rows as indicated in
shaded box in $\mathbf{x}_N,$ i.e. in our example the original time series elements $(\#9,\#5,\#1)$ represents the first point of the reconstructed phase space trajectory in $m=3$ dimensions.

Embedding parameters $m$ and $\tau$ are crucial in the analysis. Two common methods of their determination is presented in the following based on \citet{Hegger1999}.
\begin{itemize}
\item Determining time lag by mutual information. Calculating the time delayed mutual information of a time series -- similarly to autocorrelation -- one has to compute
  \begin{equation}
I(\tau)=\sum_{t=1}^{n}p(x_t,x_{t+\tau})\ln\frac{p(x_t,x_{t+\tau})}{p(x_t)p(x_{t+\tau})}
  \end{equation}
  where $p(x_t)$ is the probability of observing $x_t,$
  $p(x_t,x_{t+\tau}$ defines the joint probability distribution of
  observing $x_t$ and $x_{t+\tau}.$ Periodic time series show periodic
  mutual information function. In addition, chaotic time series also
  exhibit peaks in mutual information plot although these are not as
  strong and also not periodic. If there exists an obvious minimum of
  the  mutual information function at a certain value of $\tau,$ then
  this is a good candidate for the potential time delay. An example of
  time lag determination is shown in Figure~\ref{fig:mi}. 

\begin{center}
  \begin{figure}
  \includegraphics[angle=0,width=0.95\columnwidth]{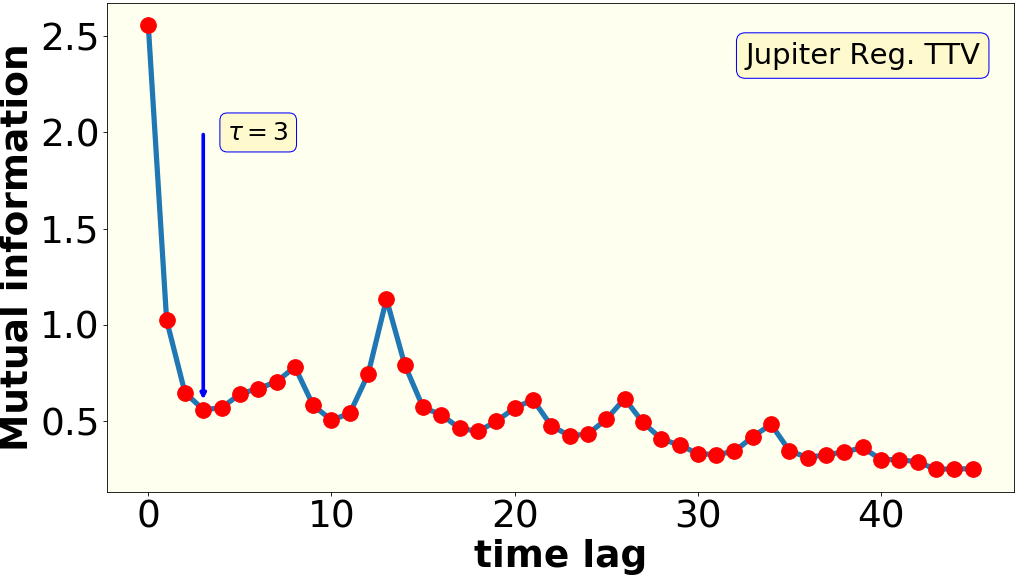}
\caption{Time delay estimation based on the first minimum of the mutual
  information function. Jupiter's TTV from regular domain
  ($(a_{\mathrm{Saturn}},e_{\mathrm{Saturn}})=(7.2,0.02)$) has been used
  as the input time series.}
\label{fig:mi}
\end{figure}
\end{center}
\item Making use of false nearest neighbours (FNN) to
    establish appropriate embedding dimension. Suppose the minimum
    required embedding dimension of a delay time series, $\mathbf{x},$
    is $m_0.$ This means that for known $\tau$ the embedded trajectory
    and the original phase space trajectory share topological
    properties. Due to the smoothness of the dynamics one also expects
    that neigbour points after one step forward iteration are mapped
    onto neighbourhoods (depending on the local dynamics,
    i.e. Lyapunov exponents). Now, let us assume the time series is
    embedded into $m<m_0$ dimensional space. As a consequence of this
    projection the topological structure is no longer preserved. That
    is, some points are projected to the neighbourhood of a point that
    would not happen in higher dimension, $m_0$. These points are
    called false neighbours. Now, if the dynamics is applied in $m$
    dimensions, these points do not remain close to each other in the
    next iteration step. They image is going to be mapped somewhere
    else resulting in a larger diameter in phase space. To quantify
    the ratio of false nearest neighbours for a trajectory the
    following expression is calculated
  \begin{equation}
    R_i=\frac{\parallel \mathbf{x}_{i+1}-\mathbf{x}_{j+1}\parallel}{\parallel \mathbf{x}_i-\mathbf{x}_j\parallel},
  \end{equation}
  where $\parallel .\parallel$ denotes Euclidean distance and
  $\mathbf{x}_j$ is nearest neighbour of $\mathbf{x}_i.$ If $R_i$
  exceeds a predefined threshold $\mathbf{x}_i$ is marked as having a
  FNN in $m$ dimension. Calculating $R_i$ for the whole time series,
  the ratio of false nearest neighbours can be obtained. The same
  procedure should be repeated for different embedding dimensions. The
  criterion of a sufficiently large embedding dimension, $m_0,$ is
  that the ratio of FNNs must be zero or close to it. In this case the
  dynamics of the original trajectory are being unfolded. Evaluation
  of embedding dimension based on FNN method is displayed in Figure~\ref{fig:fnn}.

\begin{center}
  \begin{figure}
  \includegraphics[angle=0,width=0.95\columnwidth]{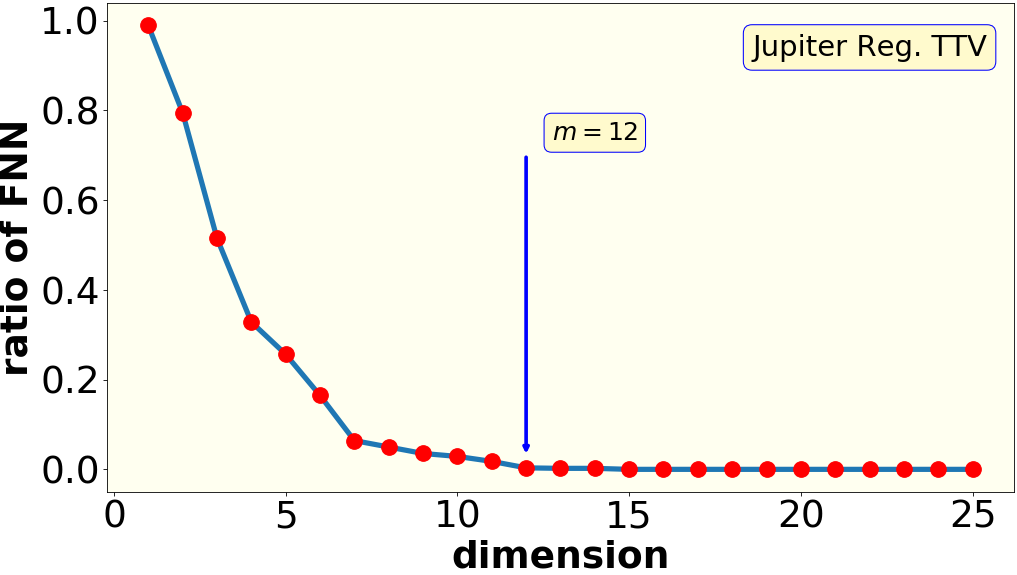}
\caption{The minimum embedding dimension requires vanishing false
  nearest neighbour ratio. The input time series is the same as in
  Figure~\ref{fig:mi}.}  
\label{fig:fnn}
\end{figure}
\end{center}
\end{itemize}

\section{Definition of various norms}
\label{app:norm}

Constructing RPs one has to choose a proper norm. Alternative norms, however, encompass different number of neighbours of a given point in the phase space. For instance, keeping the threshold $\epsilon$ fixed, $L_{\infty}$-norm finds the most, $L_1$-norm the least, and $L_2$-norm intermediate number of neighbours. \cite{Marwan2007} In what follows the definitions of different kind of norms are given. Let assume $\mathbf{p}$ and $\mathbf{q}$ are vectors in $N$ dimensional real vector space. Then the p-norm ($L_p$) of $\mathbf{p}$ and $\mathbf{q}$ reads
\begin{equation}
d_{p}(\mathbf{p},\mathbf{q})=\parallel\mathbf{p}-\mathbf{q}\parallel_{p}=\left(\sum_{i=1}^{N}(p_i-q_i)^{p}\right)^{1/p}.
\end{equation}
For different values of $p$ we get different norms:

%\begin{enumerate}
$L_1$ norm -- (Manhattan distance)
\begin{equation}
d_1(\mathbf{p},\mathbf{q})=\parallel\mathbf{p}-\mathbf{q}\parallel_1=\sum_{i=1}^{N}|p_i-q_i|.
\end{equation}

$L_2$ norm -- (Euclidean distance)
\begin{equation}
d_2(\mathbf{p},\mathbf{q})=\parallel\mathbf{p}-\mathbf{q}\parallel_2=\sqrt{\sum_{i=1}^{N}(p_i-q_i)^{2}}.
\end{equation}

$L_{\infty}$ norm -- (maximum distance)
\begin{equation}
  d_{\infty}(\mathbf{p},\mathbf{q})=\parallel\mathbf{p}-\mathbf{q}\parallel_{\infty}=\lim_{p\to\infty}\parallel\mathbf{p}-\mathbf{q}\parallel_{p}.
  \label{eq:max_norm}
\end{equation}
%\end{enumerate}
Due to its computational efficiency and analytical properties the $L_{\infty}$-norm is often used to construct recurrence plots. 

\section{Basics of networks}
\label{app:network}

In this section, only the very basics of complex networks are discussed. The purpose is to make smoother Section~\ref{sec:RN}.

Complex networks with non-trivial topology can be described by graphs, connections (edges) between their elements (vertices, nodes). In mathematical form the adjacency matrix ($\mathbf{A}$) lists the set of neighbours, i.e. those vertices that are connected. In this study recurrence networks can be described by undirected, unweighted graphs. This means connections between the nodes are bidirectional. Furthermore, the edges are equal, no weights are associated to them. Therefore, matrix $\mathbf{A}$ is symmetric. Equation~\ref{eq:adj_mat_shape} displays a fraction of matrix $\mathbf{A}$ based on TTV signal of Jupiter, see Fig.~\ref{fig:RV_TTV}e and also the associated RP in Fig.~\ref{fig:rp}.

\begin{equation}
  \mathbf{A}_{17:27,15:25}=
\begin{pmatrix}
 0& 0& 0& 0& 0& 0& 0& 1& 0& 0\\ %& 0& 0& 1& 0\\
 0& 0& 0& 0& 0& 0& 0& 0& 1& 0\\ %& 0& 0& 0& 0\\
 0& 0& 0& 0& 0& 0& 0& 0& 0& 1\\ %& 0& 0& 0& 0\\
 1& 0& 0& 0& 0& 0& 0& 0& 0& 0\\ %& 1& 0& 0& 0\\
 0& 1& 0& 0& 0& 0& 0& 0& 0& 0\\ %& 0& 1& 0& 0\\
 0& 0& 1& 0& 0& 0& 0& 0& 0& 0\\ %& 0& 0& 1& 0\\
 0& 0& 0& 1& 0& 0& 0& 0& 0& 0\\ %& 0& 0& 0& 1\\
 0& 0& 0& 0& 1& 0& 0& 0& 0& 0\\ %& 0& 0& 0& 0\\
 1& 0& 0& 0& 0& 1& 0& 0& 0& 0\\ %& 0& 0& 0& 0\\
 0& 0& 0& 0& 0& 0& 1& 0& 0& 0\\ %& 0& 0& 0& 0\\
 %0& 0& 1& 0& 0& 0& 0& 1& 0& 0& 0& 0& 0& 0\\
 %0& 0& 0& 0& 0& 0& 0& 0& 1& 0& 0& 0& 0& 0\\
 %0& 0& 0& 0& 1& 0& 0& 0& 0& 1& 0& 0& 0& 0\\
 %0& 0& 0& 0& 0& 0& 0& 0& 0& 0& 1& 0& 0& 0\\
\end{pmatrix}
\label{eq:adj_mat_shape}
\end{equation}

The non-zero entries of Equation~\ref{eq:adj_mat_shape} denote those links of the graph that are connected. One can observe that matrix $\mathbf{A}$ contains 11 neighbours out of 100. This shows the pre-defined condition of density (recurrence rate) set to 10\%.

Most of the complex networks display nontrivial structures. This feature
can be shaped into mathematical form such us degree distribution,
clustering, hierarchical structures. These measures also refer
quantitatively to the topology of RPs and, therefore, they make connections
to dynamics.

\begin{center}
  \begin{figure}
  \includegraphics[angle=0,width=0.95\columnwidth]{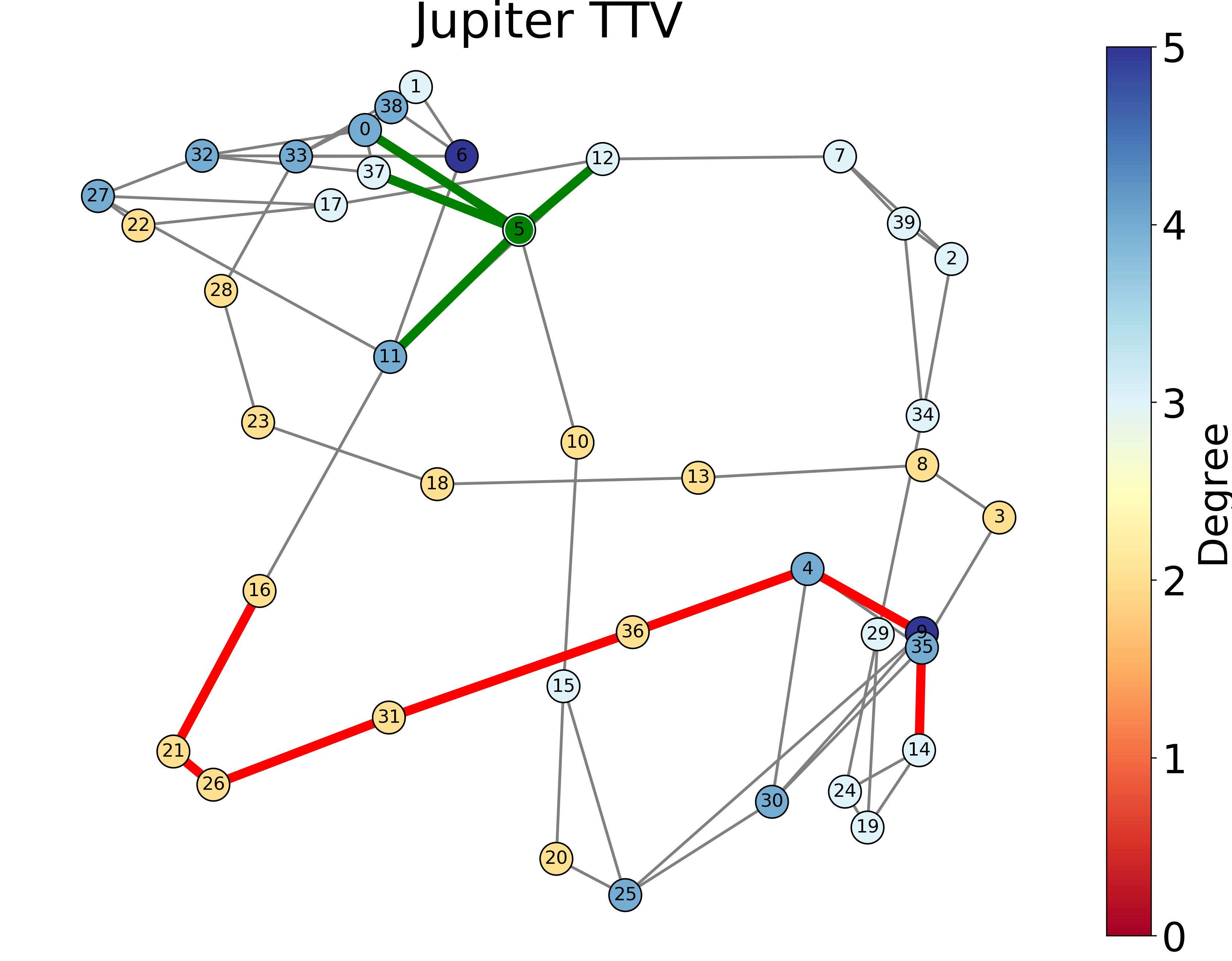}
\caption{Part of recurrence network based on Jupiter's TTV. The color bar indicates the degree of nodes, i.e. how many connections they have. Numbers in circles specify node labels.}  
\label{fig:ttv_graph}
\end{figure}
\end{center}

Now, let us take a look at the measures introduced in
Section~\ref{sec:RN}. Average path length (APL) defined by
Equation~\ref{eq:apl} is the average number of steps along the
shortest paths for all possible pairs of nodes. Geometrical
illustration can be seen in Figure~\ref{fig:ttv_graph}.
The shortest path between nodes 14 and 16 is marked by red. In order to obtain APL, one has to find all the shortest paths for all possible pairs of nodes then normalize to the number of pairs.

The other measure, transitivity (TRN), gives the probability that the adjacent
vertices are also connected. In other words, ''friends of my friends
are my friends''. TRN is calculated by the ratio between the observed
number of closed triplets and the maximum possible number of closed
triplets in the graph, Equation~\ref{eq:trn}. Transitivity is modelled
by green thin edges around vertex 5 in Figure~\ref{fig:ttv_graph}.
Node 5 has four neighbours (0,37,11,12). The question is how many triangles can be formed by these five vertices subject to node 5 has to be one of the vertices. The number of such possible triangles is six but only one (0,5,37) is actually established since nodes 0 and 37 are also neighbours. The other 'friends' of node 5 are not connected. Transitivity is the measure of these 'real' triangles compared those triplets that do not form a triangle.

Both APL and TRN refers quantitatively to the underlying network topology. Since in our case recurrence networks are strongly related to system dynamics, one can expect that various network measures (such as APL, TRN) describe the dynamical behaviour accurately as demonstrated in Sections~\ref{sec:RN} and \ref{sec:noise_gap}.  

\section{RN measures in high-dimensional maps}
\label{app:4Dstm}

In this section we investigate how the RN measures vary when the
dimension of the system is increasing. The main objective is to
demonstrate that transitivity ($TRN$) might have lower values for
ordered motion than for chaotic ones in higher dimensions. 
As a model for our study we consider the 4D and 18D symplectic maps
consisting of 2 and 9 coupled standard maps. The general formalism of
symplectic coupling of $N$ symplectic 2D maps is the following
\begin{equation}
 \left.\begin{aligned}
     \Theta^{n+1}_{i}&=\Theta^{n}_{i} +p^{n+1}_{i},\\
     p^{n+1}_{i} &=p^{n}_{i} +K \sin\Theta^{n}_{i}-\sum_{j=1}^{N}B_{i,j} \sin[\Theta^{n}_{j}-\Theta^{n}_{i}],
       \end{aligned}
 \right.
 \qquad \mod(2\pi)
 \label{eq:cstm}
\end{equation}
where $K$ is the nonlinearity parameter and $B_{i,j}$ is the coupling
strength. The coupling on the $i$-th map is a perturbation in $p_i$ and the full
coupling is symplectic provided $B_{i,j}=B_{j,i}=B.$
System~(\ref{eq:cstm}) is a typical chaotic conservative system with
mixed phase space. To show the behaviour of RN measures we fix the
parameters $K$=1.5 and $B$=0.0, 0.005, and 0.05. Furthermore, the
initial conditions are the following (${\Theta_{1}}_{0},{p_{1}}_{0}$)=(3.241,${p_1}_{0}$) where
${p_1}_{0}\in[0,2\pi],$  (${\Theta_{i}}_{0},{p_{i}}_{0}$)=(3.24,0.05),
$i$=2$\dots$9.

\begin{center}
  \begin{figure*}
  \includegraphics[angle=0,scale=.45]{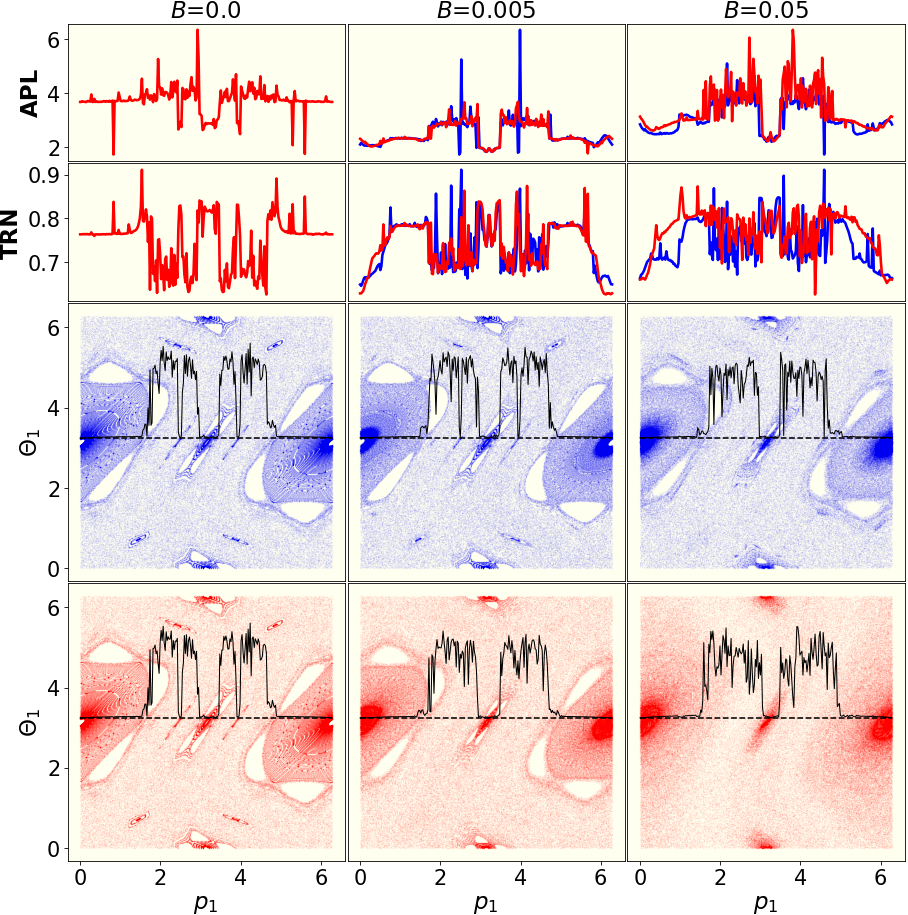}
\caption{RN measures $APL$ (top) and $TRN$ (middle) for coupled 4D
  standard map ($N$=2). The larger the $B,$ the stronger the coupling. The
  pattern specifies the lower TRN for regular motion when the extended
  phase space becomes more significant. Bottom: Phase portraits
  of ($\Theta_1,p_1$) sections. As expected the well-known embedded
  phase space structure vanishes in higher dimension. Note
  that the LCE curves (purple) does not show the real value, they are enlarged
  (by a factor of 10) for better visualization. Red part: now $N$=9, that is, the phase space has 18 dimensions.}  
\label{fig:4DK15}
\end{figure*}
\end{center}

%\begin{center}
%\begin{figure*}
% \includegraphics[angle=0,scale=.45]{./stmap_RNA_18D_K15_lowres.png}
%\caption{The same as in Figure~\ref{fig:4DK15} except that now
%  $N$=9, that is, the phase space has 18 dimensions.}  
%\label{fig:18DK15}
%\end{figure*}
%\end{center}

Phase portraits for different $B$ parameters and $N$=2 are shown in
Figure~\ref{fig:4DK15} bottom row. The classical structure is depicted when
$B$=0, i.e. the two maps are decoupled. Invariant curves fade in
higher dimension since in this case ($\Theta_{1},p_{1}$) section is a
projection of the 4D phase space. Indicating the stability we overplot
the Lyapunov exponents along the line of initial conditions. One can
observe that the main islands remain stable and only the small
stability regions become unstable for larger $B.$

Considering the recurrence network measures the 2D map serves what we
expect from literature \citep{Zou2016}. That is, smaller value of APL but larger value
of TRN for periodic motion. In turn, when the phase space is extended
the value of $TRN$ starts to decrease at regular domains. The higher the
coupling, the larger the $TRN$'s decay, $APL$ shows the same
characteristic. In other words, the large stability island keeps its
stability (the LCEs remain zero) while the network measure $TRN$ turns
to be low. Similar behaviour can be seen in \citep{Marwan2015} when
they found smaller $TRN$ for stable motion ($LCE$=0) than for unstable
in Lorenz96 model. Although, the Lorenz96 system is a high dimensional
continuous dynamical model, the explanation of the phenomenon can be
the same. Namely, in higher dimensions, periodic orbits are not
confined to a finite part of the phase space but can sweep a large
domain that is comparable with chaotic realm in size. Consequently,
the clustering, and also the transitivity, due to the divergence
of the trajectories is not so efficient in resulting in smaller values
even in the case of periodic motion. This fact might clarify the lower
TRN values for regular dynamics in the three body problem,
Section~\ref{sec:RN} Fig.~\ref{fig:rna-ae-map}.

For higher dimensional phase space, for instance $N=9,$
the tendency is similar. As one can see the
essential difference between the uncoupled and coupled scenarios
follows our former observation. The phase portrait get more fuzzy and
with this the RN measures drop off for regular motion.

Based on these findings we believe in the results presented in  Fig.~\ref{fig:rna-ae-map}(d)
wherein smaller TRNs describe the regular dynamics. The
precise study of RQA and RN description of high-dimensional
Hamiltonian systems (either continuous or discrete) is crucial and
must be carried out carefully in the future. To our best knowledge it
has not been done yet. We postpone this work
elsewhere.  

\section{Dynamic Time Warping}
\label{app:DTW}

As we have seen, the noise radius $\rho$ is a pivotal parameter in
pseudo-periodic twin surrogates algorithm (PPTS), see
Section~\ref{seq:ppts}. Its value tunes the PPTS which means if $\rho$ is
too large the generated surrogate will be a sequence of random
values. While, in contrast, if the value of $\rho$ is too small, the
produces time series is identical to the original one.
What we need, therefore, is a method that gives a suitable noise
radius which is
not too large and not too small. In other words, we want to generate a
time series that contains some dynamics noise, nevertheless, it is
similar to the original signal compared by naked eyes.

In this section we 
propose the method of dynamics time warping (DTW) that is suitable to
quantify the similarity of two time series \citep{Berndt1994}. Without
going into the details we present some of the basic feature of DTW and
then present how it works in case of PPTS algorithm.

Comparing two sequences, possible with different length, one needs a local cost measure. This
measure is typically small if the two signals are similar to each
other, and large otherwise. Then one can construct the cost matrix
that defines the local cost measure between each pair of points of two
signals. As a final step the task is to find an alignment between of
the two time series that minimizes the overall cost. For a more
precise mathematical formulation of DTW see reference \citep{Mueller2007}.

\begin{center}
\begin{figure}
  \includegraphics[angle=0,width=\columnwidth]{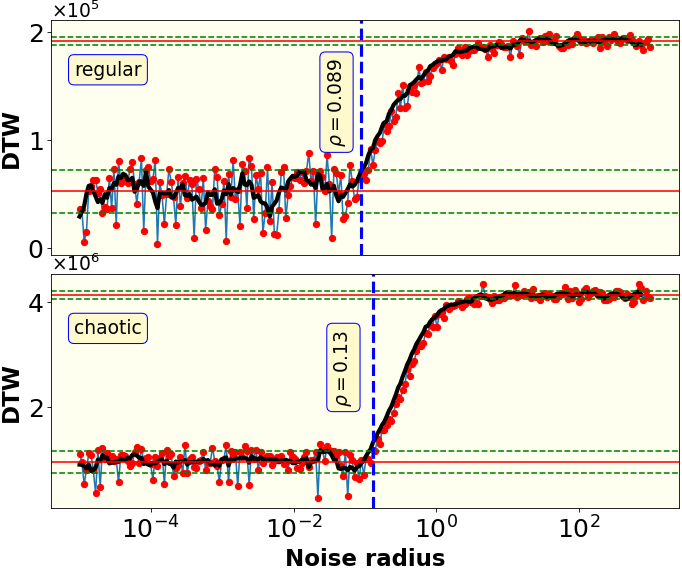}
\caption{Dynamic time warping vs. noise radius. Top: regular dynamics. Bottom: chaotic dynamics. The reference signal in both cases is Jupiter's TTV, see Fig.~\ref{fig:conf-ttv}(a) and (b). Different scale of dynamical noise in PPT surrogate algorithm yields distinct features in resulting surrogate time series. The intermediate values present a suitable noise radius to generate PPTSs having the same intracycle dynamics (but vanishing the intercycle structures) as the original data set. Red solid lines correspond to mean DTW value of first and last 20\% of data points, respectively. The green dashed lines mark the $\pm$ 1 standard deviation of the same segments of data. The thick black curve represents a smoothing of the original 250 data points by a moving average (window size is 5). }
\label{fig:dtw_rho}
\end{figure}
\end{center}

Based on the algorithm in
Section~\ref{seq:ppts} a large number (250) of PPTSs has been generated over
several orders of magnitude of $\rho.$ Then 
we stored the actual value of DTW. Small DTW characterizes minor
difference between the original signal and the generated surrogate
while relative large value reveals significant contrast between
them. See Figure~\ref{fig:dtw_rho} for specifying the noise radius in
two different dynamical regime. 

For practical purposes we accept the appropriate noise radius when the DTW curve
(black solid line) leaves the 1 standard deviation defined by the first 20\%
of data points (green dashed line). Using this rule is a must because the first part of the plot, i.e. the DTW for small noise radii, might fluctuate in a great way, especially in strongly chaotic cases.

The upper panel of Fig.~\ref{fig:dtw_rho} portrays the relationship between $DTW$ and noise radius for the regular orbit originating at $(a_{\text{Saturn}},e_{\text{Saturn}})=(7.2,0.02).$ The two well separated values of $DTW$ indicate the role of the noise radius explained above. The smaller value of $DTW$ ($\rho\lesssim 0.1$) characterizes that the two time series are similar while the right tale of the plot ($\rho\gtrsim 3$) denotes when the surrogate and the original signal are completely different. The noise radius should be chosen from the rest part of the curve. The blue dashed line is placed to the value of $\rho=0.075$ where $DTW$ starts to diverge from its mean value ($\sim 5\times 10^4$) calculated from the first 20\% of the data points.

The lower panel shows $DTW$ vs. noise radius for the chaotic orbit starting at ($a_{\text{Saturn}},e_{\text{Saturn}})=(8.0,0.2$). The main structure of the plot matches the upper one. However, the $DTW$ values are larger by an order of magnitude. The noise radius corresponding to the 1 standard deviation limit ($\rho=0.13$) is marked by the vertical blue dashed line.

Making use of DTW we are able to set the appropriate value of noise radius ($\rho$) in PPTS algorithm, see Eq.~\ref{eq:noise_radius}. We should however emphasize that the choice of 1 sigma limit is empirical and further investigation is needed how to suit more precisely the DTW algorithm to find the best $\rho.$
%%%%%%%%%%%%%%%%%%%%%%%%%%%%%%%%%%%%%%%%%%%%%%%%%%

% Don't change these lines
\bsp	% typesetting comment
\label{lastpage}
\end{document}